\documentclass[twocolumn,aps,pra,showpacs,longbibliography,10pt]{revtex4-1} 

\usepackage[normalem]{ulem}
\usepackage{amsfonts,amssymb,amsmath,graphicx,color}
\usepackage{array,multirow}
\usepackage{enumitem}
\usepackage{braket}
\usepackage[version=4]{mhchem}
\usepackage{float}
\usepackage[colorlinks=true,linkcolor=blue,citecolor=blue,urlcolor=blue]{hyperref}

\newcommand{\red}[1]{{\textcolor{red}{#1}}}
\newcommand{\ie}{{\it i.e.~}} 	
\newcommand{\eg}{{\it e.g.~}} 	
\newcommand{\ci}[0]{\mathrm{i}} 
\newcommand\commentout[1]{}

\newcolumntype{P}[1]{>{\centering\arraybackslash}p{#1}}

\definecolor{greenPR}{rgb}{0.00, 0.6, 0.00}

\begin{document}

\title{Theory of magnetotransport in shaped topological insulator nanowires}
\author{Ansgar Graf$^{1,2}$}
\author{Raphael Kozlovsky$^1$}
\author{Klaus Richter$^1$}
\author{Cosimo Gorini$^1$}%
\affiliation{%
	$^1$Institut f\"ur Theoretische Physik, Universit\"at Regensburg, 93040 Regensburg, Germany\\
	$^2$Universit\'e Paris-Saclay, CNRS, Laboratoire de Physique des Solides, 91405, Orsay, France.
}%
\date{\today}

\begin{abstract}
It is demonstrated that shaped topological insulator (TI) nanowires,
\ie such that their cross-section radius varies along the wire length, can be tuned into a number of different transport
regimes when immersed in a {\it homogeneous} coaxial magnetic field.  This is in contrast with widely studied
tubular nanowires with constant cross-section, and is due to magnetic confinement of Dirac surface carriers.
In flat 2D systems, such a confinement requires inhomogeneous magnetic fields, while for shaped nanowires of standard size
homogeneous fields of the order of $B\sim\,1$T are sufficient.
 We put recent work [\citeauthor{Kozlovsky2020}, Phys. Rev. Lett. 124, 126804 (2020)]
into broader context and extend it to deal with axially symmetric wire geometries with arbitrary radial profile.
A dumbbell-shaped TI nanowire is used as a paradigmatic example for transport through a constriction
and shown to be tunable into five different
transport regimes: (i) conductance steps, (ii) resonant transmission, (iii) current suppression,
 (iv) Coulomb blockade, and (v) transport through a triple quantum dot.  Switching between regimes is achieved by
modulating the strength
of a coaxial magnetic field and does not require strict axial symmetry of the wire cross section.
As such, it should be observable in TI nanowires fabricated with available experimental techniques. 
\end{abstract}

\maketitle

\section{Introduction}

Topological materials have been a central topic in solid state research for roughly
two decades.  Many distinct topological phases are currently known \cite{bansil2016},
that of (strong) topological insulators (TIs) being a most prominent one \cite{hasan2010,ando2013}.
The low-energy electronic structure of a flat TI surface is characterized by a single Dirac cone \cite{SCzhang2009}.
This case, possibly the simplest and most widely studied one, 
is already enough to produce a number of notable transport phenomena \footnote{See \eg 
Refs.~\cite{tokoyama2010,schwab2011,lee2015} or Refs.~\cite{culcer2012,ando2017} for reviews.}.
Geometrically more complex than a flat surface, TI nanowires (TINWs) have also been intensively studied 
\cite{zhang2009, egger2010, imura2011, dufouleur2013, tian2013}.
One notable reason for this is that their high surface-to-volume ratio enhances the visibility 
of surface transport features.  Moreover, transport takes place on a surface which is closed along the transversal
direction, enclosing the (nominally) insulating three-dimensional (3D) TI bulk.  This leads to the interplay between
the spin Berry phase of surface states and an Aharonov-Bohm phase acquired in the presence 
of a coaxial magnetic field \cite{Kozlovsky2020,peng2010, ostrovsky2010, bardarson2010, zhang2010, rosenberg2010, bardarson2013, cho2015, ziegler2018}.

\begin{figure}
	\includegraphics[width=.95\columnwidth]{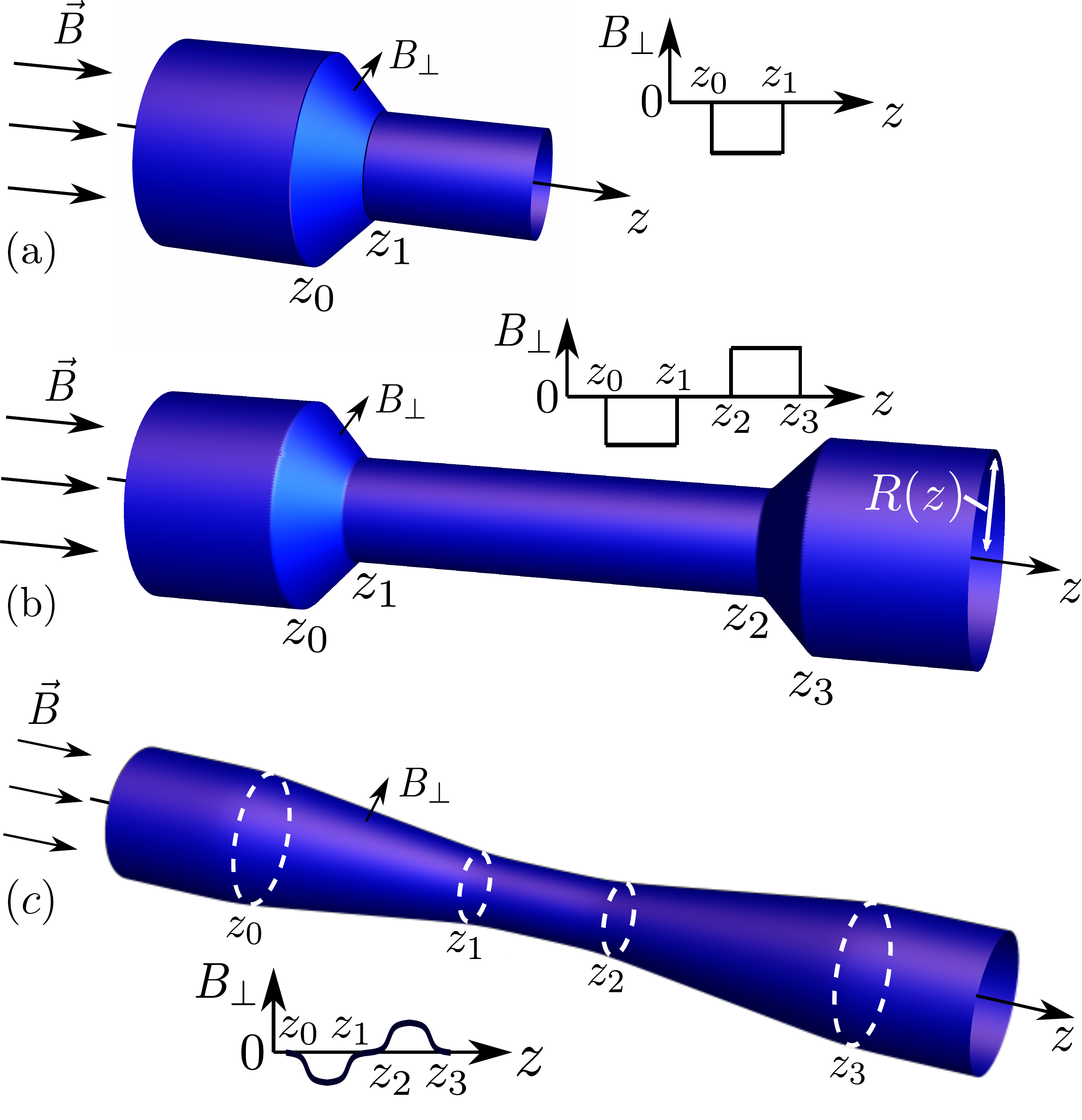}
	\caption{Examples for shaped TINWs subject to a coaxial magnetic field. The bulk is assumed to be perfectly insulating, while the electronic structure on the metallic surface (blue) is modeled by the Hamiltonian (\ref{mainham}). (a) The region between $z_0$ and $z_1$ (which correspond to radii $R_0$ and $R_1$) will be referred to as a TI nanocone (TINC). Defining $\delta\equiv\arctan\mathcal{S}$, where $\mathcal{S}\equiv(R_1-R_0)/(z_1-z_0)$ is the slope of the TINC, we get the magnetic field component $B_\perp=B\sin\delta$ piercing the surface. It is negative for $R_1<R_0$. (b) The region between $z_0$ and $z_3$ will be referred to as a TI dumbbell. For arguments concerning transport, we will consider the cylindrical leads to either side of the junctions to be metallic.
	(c) Smoothed TI dumbbell.}
	\label{fig_cones}
\end{figure}

Insight into the physics of 3DTI surfaces of more complex geometry (beyond flat or cylindrical)
can be obtained from an effective surface Dirac theory derived either from the 3D bulk Hamiltonian of the paradigmatic bismuth-based TIs 
\cite{imura2012,takane2013}, or from a field theoretic approach \cite{xypakis2017}. 
In this paper we apply such a theory to axially symmetric TINWs whose radius varies arbitrarily along their length,
which we dub from now on simply \textit{shaped TINWs}.  Our goal is a systematic study of their magnetotransport properties,
thereby extending previous work \cite{Kozlovsky2020} on truncated TI nanocones (TINCs),
see Fig.~\ref{fig_cones}(a).  The latter were shown to offer rich magnetotransport signatures, ranging from conductance quantization to resonant transmission through Dirac Landau levels and Coulomb blockade-type transport.
We will discuss how further regimes become available in shaped TINWs of experimentally realistic sizes.
Note that such TINWs are structurally shaped on mesoscopic scales, 
in stark contrast to the overall cylindrical but (randomly) rippled TINWs considered in Ref.~[\onlinecite{xypakis2017}].

When a shaped TINW is subjected to an arbitrary magnetic field,
the latter can be decomposed into two components, one perpendicular to the nanowire axis, the other coaxial. 
It is instructive to study each component separately. 
If the component perpendicular to the TINW axis is strong,
\ie the associated magnetic length $l_B$ is much smaller than the nanowire diameter,
the nanowire conductance will be dominated by chiral side (hinge) 
states \cite{Kozlovsky2020,zhang2011,vafek2011,sitte2012,brey2014,koenig2014}. These states
are largely independent of the TINW shape. This geometry-insensitive configuration is not interesting
for our purposes, ergo we focus here on purely coaxial magnetic fields, $\mathbf{B}=B\hat{z}$,
with $\hat{z}$ the coaxial unit vector.
In such a configuration the TINW transport properties depend strongly on its radial profile.
Indeed, in a shaped TINW, the magnetic flux through the nanowire cross section is a function of $z$,
and so is the out-of-surface component $B_\perp=B_\perp(z)$ experienced by the surface electrons, see Fig.~\ref{fig_cones}.
If the phase coherence length is sufficiently long, the $z$-dependent cross section leads to a 
$z$-dependent Aharonov-Bohm phase. 
We will show that the latter, together with quantum confinement due to the finite circumference, generates a mass-like $z$-dependent potential. 
This potential can be used to qualitatively predict the transport characteristics of any shaped TINW, and was recently discussed for the simplest case of a single TINC \cite{Kozlovsky2020}.
In order to understand magnetotransport in shaped nanowires, an essential step will be to consider such TINCs as building blocks of more complex geometries, see Fig.~\ref{fig_cones}.

The paper consists of three main parts, Secs.~\ref{ham}-\ref{dumbbell}, plus a concluding one, Sec.~\ref{conclusion}.    
In Sec.~\ref{ham} we derive an effective 1D Dirac equation that can be used to describe the electronic surface structure of shaped TINWs.  Such a Dirac equation can be solved analytically for simple cases and numerically for more complex TINW geometries.  The reader not interested in the technical aspect of the derivation can skip directly to
Sec.~\ref{subsec_physical_picture}, which discusses the ensuing physical picture
-- namely, that transport properties are determined by an effective mass-like potential entering the 1D Dirac equation.

Sec.~\ref{TINC} reviews pedagogically the physics of TINCs in a coaxial magnetic field.  
This is instructive, as any shaped nanowire can be constructed from a succession of infinitesimal conical segments with constant slope. 
We will show that provided $B_\perp$ is strong enough, such that $l_B$ is small compared to the length of the TINC, the conductance is determined by resonant transmission through Dirac Landau levels (LLs) that form on the wire's surface.

In Sec.~\ref{dumbbell} we introduce the dumbbell-shaped TINW of Fig.~\ref{fig_cones}(b),
representing the important case of a nanowire constriction and
simply referred to from now on as {\it TI dumbbell}.  In magnetic fields of intermediate strength ($B\sim1\,$T) such an object can be tuned into three fundamentally different transport regimes by varying the field magnitude.
First, if the lead Fermi level $E_F$ matches the energy of a LL from the side TINCs, the latter become transparent and resonant transmission into and out of the central region takes place.
If on the other hand $E_F$ lies in between LLs, the latter become opaque and act as barriers, suppressing overall transmission ($G\approx 0$).  Finally, between these two special cases, the conductance of the individual TINCs fulfills the condition $0<G\ll e^2/h$, a prerequisite for Coulomb blockade physics. We show that the Coulomb blockade regime should be accessible in the TI dumbbell, with single-particle energy levels of the confined Dirac electrons modulating the periodicity of Coulomb blockade oscillations. 
Most notably, the dependence of the transport regimes on $E_F$ allows to switch on and off Coulomb blockade physics by a simple tuning of the magnetic field strength, which shifts the LL ladder.  

We also treat the more realistic scenario of smoothly shaped TINWs, whose geometry is comparable to experimentally realized nanowires \cite{Kessel_2017}, see Fig.~\ref{fig_cones}(c).  
For such geometries, yet another transport regime emerges for high magnetic fields from the interplay between the magnetic length and the length scale of the smoothing.

Sec.~\ref{conclusion} concludes and sums up our findings.
A series of technical details are discussed in the Appendices \ref{AppA}-\ref{Appsmoothcone}.

\section{Dirac surface theory for a shaped TI nanowire}
\label{ham}
	

\subsection{Surface Dirac equation}

We are interested in topological surface transport \cite{ostrovsky2010,bardarson2010,zhang2010}, so the starting point is the surface Dirac Hamiltonian $H$. In experimental samples, the bulk is usually not perfectly insulating, but several techniques, \eg gating or compensation, can be used to suppress its transport contribution, see for example Refs. \cite{Kong_2011, ZhangJ_2011, Zhou_2012}. We thus neglect bulk contributions throughout.

The model Hamiltonian $H$ satisfies the time-independent surface Dirac equation
\begin{equation}
H\psi=\epsilon\psi
\label{surfdir}
\end{equation}
and can be derived starting from either of the two approaches mentioned above: microscopic or field theoretic. Both derivations can be consulted in detail in Ref.~[\onlinecite{xypakis2017}]. In order to introduce our notation, the derivation of $H$ is sketched very briefly in Appendix~\ref{AppA}. 

One finds
\begin{equation}
\begin{aligned}
H=v_F&\left[\frac{1}{\sqrt{1+R'^2}}\left(p_z-\frac{i\hbar}{2}\frac{R'}{R}\right)\sigma_z\right.\\&\left.+\left(p_\varphi+\frac{\hbar}{R}\frac{\Phi}{\Phi_0}\right)\sigma_y\right],
\label{mainham}
\end{aligned}
\end{equation}
where $R\equiv R(z)$ is the radius of the shaped TINW as a function of the coaxial coordinate $z$, $R' \equiv \mathrm{d}R/\mathrm{d}z$, $\Phi\equiv \pi BR^2$ the magnetic flux enclosed by the wire, $\Phi_0\equiv h/e$ the flux quantum, and $\sigma_{y,z}$ are Pauli matrices. The momentum operators are defined as $p_z\equiv -i\hbar\partial_z$, and $p_\varphi\equiv-i\hbar R^{-1}\partial_\varphi$. Note the different origin of the second and fourth terms: The shift in coaxial momentum is due to the spin connection \cite{fecko2006,koke2016}, while the shift in azimuthal momentum is due to the magnetic field. 

Note that the spin connection term in Eq. (\ref{mainham}) can be gauged away by the local transformation 
\begin{equation}
\begin{aligned}
\psi&\rightarrow\tilde{\psi}=\sqrt{R}\psi,\\
H&\rightarrow\tilde{H}=\sqrt{R}H\frac{1}{\sqrt{R}},
\label{gauge}
\end{aligned}
\end{equation}
such that
\begin{equation}
\label{almost_ourHam}
\tilde{H}=v_F\left[\frac{1}{\sqrt{1+R'^2}}p_z\sigma_z+\left(p_\varphi+\frac{\hbar}{R}\frac{\Phi}{\Phi_0}\right)\sigma_y\right].
\end{equation}
In terms of the arc length coordinate $s$ running along the TINW surface, such that
$ds^2=dz^2+dR^2$, Eq.~\eqref{almost_ourHam} becomes
\begin{equation}
\begin{aligned}
\tilde{H}=v_F\left[p_s\sigma_z+\left(p_\varphi+\frac{\hbar}{R}\frac{\Phi}{\Phi_0}\right)\sigma_y\right],
\label{ourHam}
\end{aligned}
\end{equation}
where $p_s\equiv-i\hbar\partial_s$. This last form of the Hamiltonian is particularly well suited
for simulating transport through shaped TINWs with a numerical tight-binding approach, 
see Ref.~[\onlinecite{Kozlovsky2020}] or Appendix~\ref{AppConductanceKwant} for details.

For our analytics and general discussions we will, however, express everything in terms of the coaxial coordinate $z$
throughout the paper. That is, we solve the eigenvalue problem Eq.~\eqref{surfdir} using the Hamiltonian \eqref{mainham}.
Exploiting rotational symmetry, the solution to Eq.~\eqref{surfdir} can be written as
\begin{equation}
\psi=e^{-i(l+1/2)\varphi}\chi_{nl}(z),
\label{ansatz}
\end{equation}
where $\chi_{nl}(z)$ is a two-spinor and $l\in\mathbb{Z}$ denotes the orbital angular momentum quantum number. The latter can only assume discrete values due to the azimuthal size confinement. From now on $\chi_{nl}\equiv\chi_{nl}(z)$ for brevity. 
The shift of $1/2$ in the angular momentum quantization represents the presence of a spin Berry phase of $\pi$, which is a distinct feature of 3D TINWs \cite{zhang2009,bardarson2010,zhang2010,imura2011}. The spin Berry phase ensures antiperiodic boundary conditions in the azimuthal direction. The meaning of the quantum number $n\in\mathbb{N}$ will become clear in Sec.~\ref{TINC}. (Essentially, for a given angular momentum $l$, it labels a series of bound, quasi-bound and/or scattering states depending on the character of the corresponding effective potential landscape, cf. Eq. (\ref{effpot}) below.) With the ansatz (\ref{ansatz}), we obtain the 1D Dirac equation
\begin{equation}
\left[\frac{v_F}{\sqrt{1+R'^2}}\left(p_z-\frac{i\hbar}{2}\frac{R'}{R}\right)\sigma_z+V_l\sigma_y\right]\chi_{nl}=\epsilon_{nl}\chi_{nl}.
\label{Direq}
\end{equation}
Here, the angular momentum term
\begin{equation}
V_l\equiv\hbar v_F k_l\equiv-\frac{\hbar v_F}{R}\left(l+\frac{1}{2}-\frac{\Phi}{\Phi_0}\right)
\label{effpot}
\end{equation}
induces a position- and magnetic field-dependent energy gap. For given $B$, this leads to a mass-like potential landscape along the wire that, unlike an electrostatic potential, does not admit Klein tunneling \cite{allain2011}. Consequently, the sign of $V_l$ is unimportant: a state with angular momentum quantum number $l$ sees the \textit{effective potential} $|V_l|$.

The role of $|V_l|$ will be demystified in Sec.~\ref{subsec_physical_picture}. Here we just add two remarks. 
(i) It is enlightening to consider the limit of a cylindrical TINW. In this case, $|V_l|$ is constant along the wire and simply equal to the energy minimum of the corresponding subband $\epsilon_{l}(k_z)$. The quantum number $n$ then takes the form of a continuous coaxial wave number: $n\to k_z$. (ii) In Sec.~\ref{TINCpot}, we will discuss in detail the effective potential for a TINC. In the course of this, we will point out (and elaborate on it in Appendix~\ref{AppB}) that mass potential landscapes analogous to $|V_l|$ appear in any system where Dirac carriers feel an effectively inhomogeneous magnetic field, for example when magnetic step barriers are formed in graphene \cite{demartino2007,masir2008}.

\subsection{Solution of the surface Dirac equation}
\label{numerics}

Equation (\ref{Direq}) can be decoupled into two second-order partial differential equations. We define $\gamma\equiv1/\sqrt{1+R'^2}$
for compact notation, insert the two-spinor $\smash{\chi_{nl}=(\chi_{nl}^{(1)},\chi_{nl}^{(2)})^T}$ and define $\smash{\chi_{nl}^\pm\equiv\chi_{nl}^{(1)}\pm\chi_{nl}^{(2)}}$, such that the Dirac equation (\ref{Direq}) becomes
\begin{equation}
	\mathcal{O}_l^\pm\chi_{nl}^\pm=\epsilon_{nl}^2\chi_{nl}^\pm,
\label{laplaceeqq}
\end{equation}
with
\begin{equation}
\mathcal{O}_l^\pm\equiv-(\hbar v_F\gamma)^2\left[\partial_z^2-R'\left(\gamma^2R''-\frac{1}{R}\right)\partial_z+\mathcal{P}^\pm_l\right].
\end{equation}
The magnetic field and angular momentum dependence enter only into the last term,
\begin{equation}
\begin{aligned}
\mathcal{P}^\pm_l&\equiv-\frac{1}{\gamma}\left(\frac{k_l^2}{\gamma}\mp k_l'\right)+\frac{1}{2R}\left(\gamma^2R''-\frac{R'^2}{2R}\right).
\end{aligned}
\end{equation}
Let us check the limit of a cylindrical TINW: $R'=k_l'=0$, $\gamma=1$, $\smash{\chi_{nl}^\pm\rightarrow e^{ik_zz}(\chi_l^{(1)}\pm\chi_l^{(2)})}$ and $\epsilon_{nl}\rightarrow\epsilon_{l}(k_z)$, where $\smash{\chi_l^{(1,2)}}$ are independent of $z$. This yields the energy dispersion $\epsilon_{l}(k_z)=\pm\hbar v_F\sqrt{k_z^2+k_l^2}$, as expected \cite{bardarson2013,ziegler2018}.

For an arbitrary geometry $R(z)$, magnetic field $B$ and angular momentum quantum number $l$, one can numerically solve Eq.\ (\ref{laplaceeqq}). The numerical implementation of $R(z)$ is described in Appendix~\ref{Appsmoothcone}. 

\subsection{Effective mass-potential -- physical picture}
\label{subsec_physical_picture}
\begin{figure}
	\centering
	\includegraphics[width=\columnwidth]{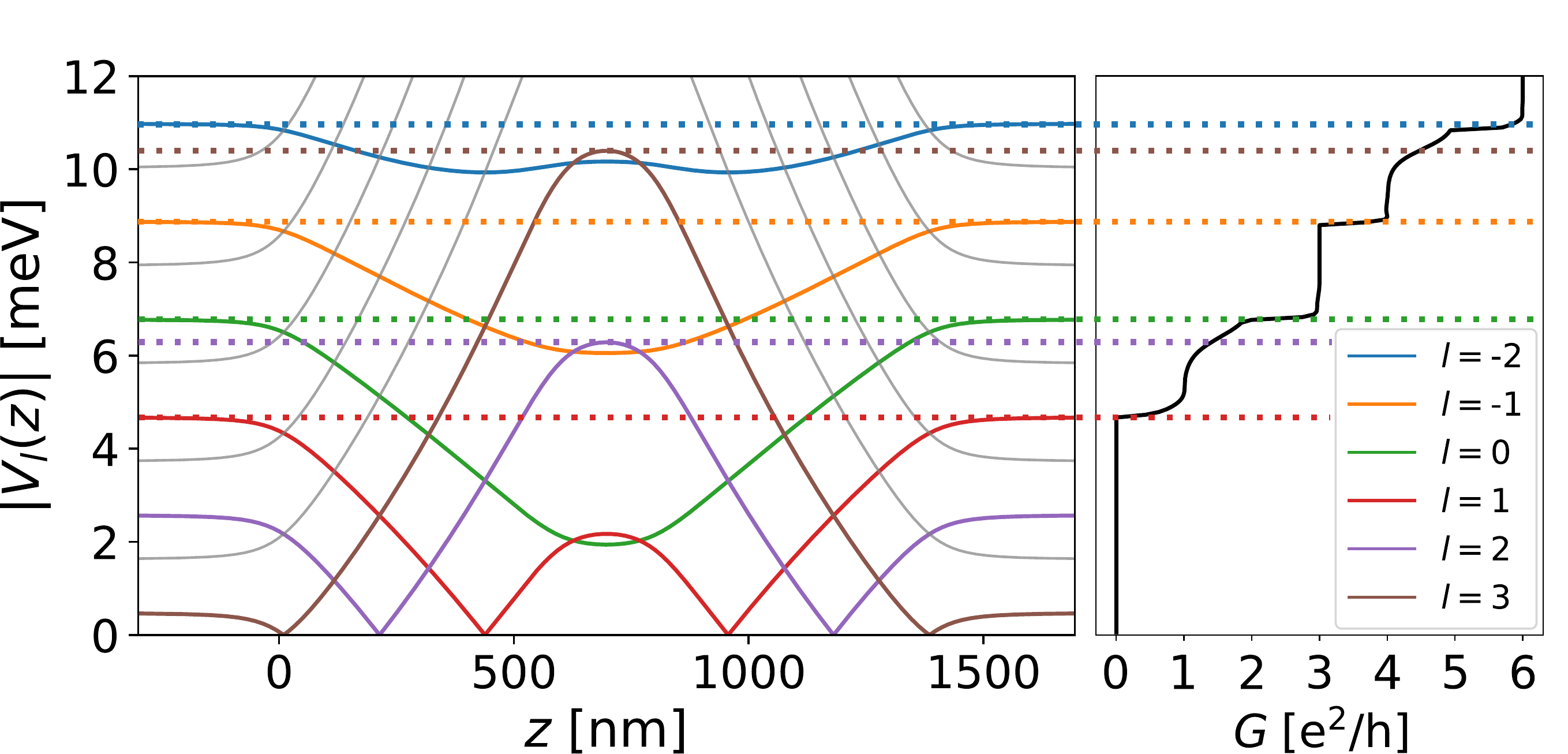}
	\caption{Left panel: Effective mass potential along a smoothed dumbbell in a magnetic field of $B=200\,$mT. Right panel: Conductance (horizontal axis) as a function of the Fermi energy (vertical axis). The scale is the same as in the left panel. Parameters of the smoothed dumbbell: $R_0 = 156.6\,$nm, $R_1 = R_0/2$, $z_1 = 594.7\,$nm, $z_2=800\,$nm, $\sigma=10\,\mu$m$^{-1}$.
}
	\label{G_smoothedDB}
\end{figure}
The purpose of this Subsection is to convey a better understanding of the role of the effective mass-potential $|V_l|$, Eq. (\ref{effpot}). We therefore visualize its effect on the transport properties of a shaped TINW in a coaxial magnetic field by presenting a concrete example, namely a smoothed TI dumbbell [cf. Fig.~\ref{fig_cones}(c), the corresponding radial profile being defined in Appendix~\ref{Appsmoothcone}] with parameters given in Fig. \ref{G_smoothedDB}.

Note that, in Section \ref{dumbbell}, we will discuss the magnetotransport properties of such TI dumbbells in great detail, for all magnetic fields from $B=0$ to several Tesla, and the choice of parameters will be very relevant; in particular, we will mostly focus on the regime of about one Tesla, where Coulomb blockade physics arises. In the present Subsection, in contrast, our aim is purely pedagogical, so the dimensions of the junction and $B$-value are not overly important. In particular, the low value $B=200\,$mT chosen here is pedagogically useful (not too many $l$-modes present in Fig. \ref{G_smoothedDB}), but not a very interesting choice from the point of view of Section \ref{dumbbell}.

In order to analyze the pristine effect of $|V_l|$ on the conductance of the dumbbell, we take a clean system (no disorder) with simple cylindrical TINW leads attached.
The conductance simulations are performed with \textsc{kwant} \cite{groth2014}. For details on the numerics, such as the non-uniform lattice we use, we refer to Appendix~\ref{AppConductanceKwant}.  Note that we fix the Fermi velocity to $v_F=5\cdot10^5\,$m$\,$s$^{-1}$ for all our numerical calculations, which is a typical value for bulk TIs \cite{SCzhang2009}.

The effective potential $|V_l(z)|$ for the set of parameters given in the caption of Fig.~\ref{G_smoothedDB}, and the corresponding conductance $G$ as a function of the Fermi level $E_F$, are displayed in Fig.~\ref{G_smoothedDB}. The transport characteristics of the smoothed TI dumbbell can be qualitatively understood in terms of $|V_l|$ in the following way: Transport is blocked below $E_F\approx 4.7\,$meV (below the red dashed line) because all open
lead modes [\ie modes with $\epsilon_l(k_z=0)<4.7\,$meV in the cylindrical leads, see $l = 3$ (brown) and $l = 2$ (purple)] do not have enough energy to overcome the central barriers.
For $E_F \gtrsim 4.7\,$meV, the $l = 1$ mode can pass through the TINW, since the
local maximum of $|V_{l=1}|$ (red) at the wire center is only slightly above 2$\,$meV. 
Hence we observe a conductance step at $E_F\approx 4.7\,$meV.
The situation is slightly different for the $l = 2$ mode (purple). 
It is already present in the leads at $E_F\approx 2.6\,$meV, but the potential has its maximum in the center at
$|V_{l=2}|\approx 6.3\,$meV. 
Comparing the corresponding conductance step with the one at $E_F\approx 4.7\,$meV,
it is apparent that its slope is lower.
The reason for this is that the electronic mode can tunnel through the $l = 2$ barrier at energies below $6.3\,$meV, leading to a finite conductance contribution at lower energies.
The same behavior can be observed for the $l=3$ mode. This exemplary discussion shows that plotting $|V_l|$ for all values of $l$ relevant at low energy is an efficient way to predict qualitative features of the conductance $G(E_F)$ in shaped TINWs, for any given $B$.

Note that due to rotational symmetry, coupling between different $l$-modes is
absent and the crossings in  Fig.~\ref{G_smoothedDB} are real crossings, not avoided crossings.
Thus an electron cannot traverse the wire by changing its orbital angular momentum
quantum number. 
Rotational symmetry is broken for instance by disorder, which allows coupling between $l$-modes. 
We will see the effect of $l$-mode coupling later on.

\section{Magnetotransport characteristics of a TI nanocone}
\label{TINC}

The TI nanocone (TINC) depicted in Fig.~\ref{fig_cones}(a) represents  
the elementary building block of shaped TINWs, so we shall study in detail its transport characteristics.
The latter were recently pointed out in Ref.~[\onlinecite{Kozlovsky2020}], and we provide here
a more pedagogical and complete treatment of the subject, focusing exclusively on coaxial magnetic fields.

\subsection{Effective mass-potential of a TI nanocone}
\label{TINCpot}
\begin{figure}[h!]
	\centering
	\includegraphics[width=.98\columnwidth]{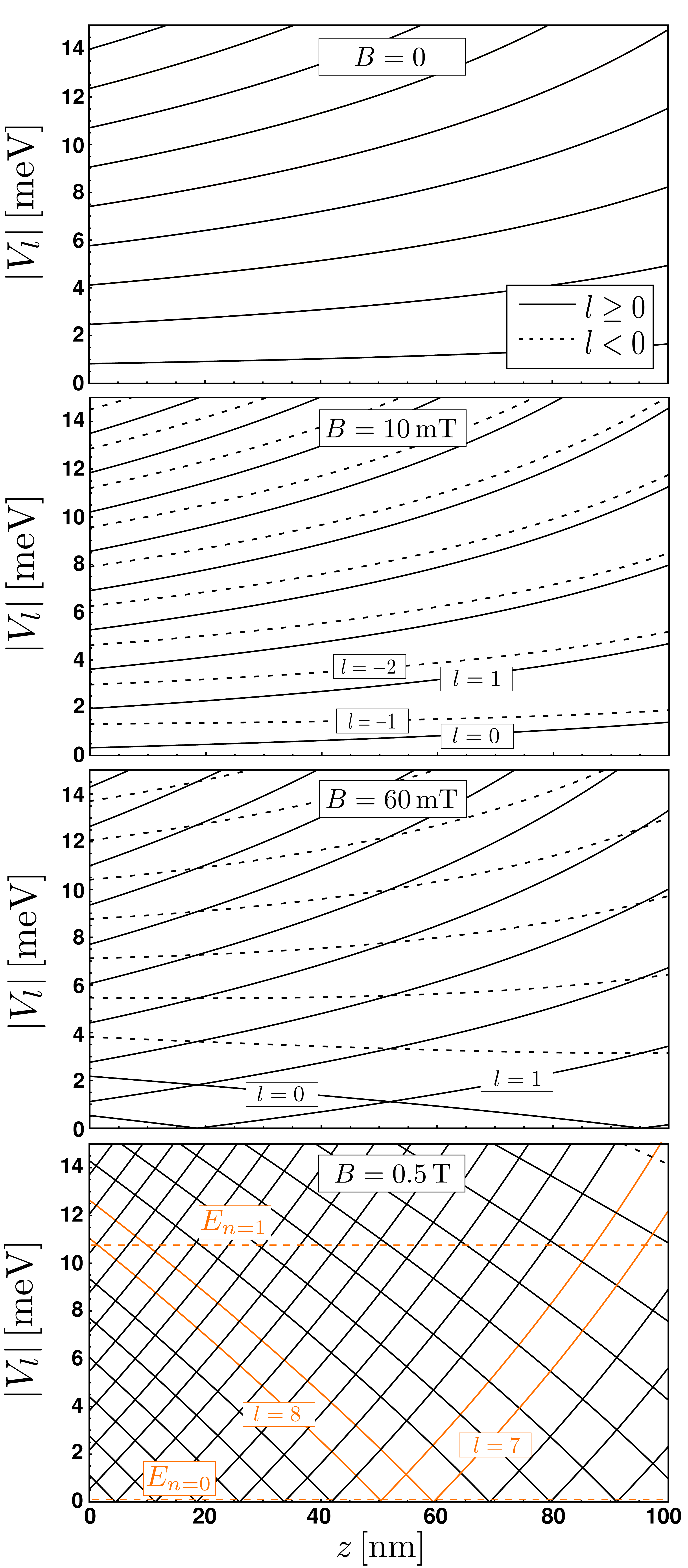}
	\caption{Effective potential landscape $|V_l(z)|$ on a TINC as depicted in Fig.~\ref{fig_cones}(a) with $z_0=0$, $z_1=100\,$nm, $R_0=2R_1=200\,$nm,  for several magnetic field strengths and as obtained from Eq.~(\ref{effpot}).
	Upon increasing $B$, modes with $l\geq0$ become dominant at low energies, and a growing number of them features a root on the TINC. Around these roots, wedge-shaped potential wells are formed, giving rise to bound states at Dirac LL energies $E_n$, cf. Eq. (\ref{LLens}). 
    }
	\label{fig:nanocone}
\end{figure}
For a given TINC geometry, Fig.~\ref{fig:nanocone} shows how the potential $|V_l|$ evolves 
as a function of the magnetic field.
For $B=0$, $|V_l|\propto1/R$ due to size confinement only, and modes with angular momentum quantum number $l$ and $-l-1$ are degenerate.  
Upon turning on the magnetic field the degeneracy is lifted: 
modes with $l<0$ move upward in energy, while $l\geq0$ modes move downward.  The potential for a given mode possesses a root
inside the TINC as soon as the critical radius
\begin{equation}
	\tilde{R}_l(B)\equiv\sqrt{\frac{\Phi_0}{\pi B}\left(l+\frac{1}{2}\right)}
	\label{critrad}
\end{equation}
fulfills the condition $\tilde{R}_l(B)\leq R_0$. This root is located at
\begin{equation}
\tilde{z}_l(B)\equiv\frac{1}{\mathcal{S}}\left[\tilde{R}_l(B)-R_0\right],
\label{root}
\end{equation}
since $R=R_0+\mathcal{S}z$ with the slope
$\mathcal{S}\equiv\frac{R_1-R_0}{z_1-z_0}<0$.
The higher the magnetic field, the larger the number of effective potentials $|V_{l\geq0}|$ developing a wedge-shaped well, whose minimum migrates from $\tilde{z}_l=z_0$ to $\tilde{z}_l=z_1$ and vanishes as soon as $\tilde{R}_l(B)<R_1$.  
We will see in Sec.~\ref{TINCstruc} that this can be understood as the formation of Landau levels \footnote{
As can be seen from Fig.~\ref{fig:nanocone}, the twofold angular momentum degeneracy gets lifted for $B\neq0$ and is never restored for higher $B$, due to the out-of-surface component. This is in marked contrast to cylindrical TINWs.}.

Note the interesting analogy to graphene in an inhomogeneous magnetic field:
Consider the step-like profile of $B_\perp$, see inset to Fig.~\ref{fig_cones}(a). Due to the Dirac character of the surface electrons, we expect similar physics as for a magnetic step barrier in graphene, studied in Ref. \cite{demartino2007}. 
In the same way, more complicated profiles of $B_\perp$, see for example Fig. \ref{fig_cones}(b), are analogous to more complicated magnetic barriers in graphene, see Ref. \cite{masir2008}. Indeed, a shaped TINW with profile $B_\perp(z)$ is, in many respects, similar to graphene subject to an equivalent inhomogeneous magnetic field. In such 2D Dirac systems, the role of the (angular momentum-based) effective potential (\ref{effpot}) is taken by a transverse momentum-based effective potential, as we point out in Appendix~\ref{AppB}; this potential develops wedges similar to those in Fig. \ref{fig:nanocone} for suitable system parameters. 

In view of this analogy (for more details see Appendix \ref{AppB}), 
one expects that a TINC may act as a strong magnetic barrier -- indeed, we will see that it does -- and that Dirac electrons can be confined in between two TINCs. One should, however, remember that a magnetic step barrier in (2D) graphene requires an \textit{inhomogenoeus} magnetic field, while the (3D) TINC allows for similar physics by just using a \textit{homogenoeus} magnetic field. 

\subsection{Electronic structure of a TI nanocone}
\label{TINCstruc}

We now discuss solutions of the 1D effective Dirac equation \eqref{laplaceeqq} for the TINC geometry with leads as depicted in Fig.~\ref{fig_cones}(a). 
Note that $|V_l|$ is constant in the leads with values $|V_l(z_0)|$ ($|V_l(z_1)|$) in the left (right) lead.
There are different regimes separated by the energy thresholds $\epsilon_l^\text{min}\equiv\text{min}(|V_l(z_0)|,|V_l(z_1)|)$ and $\epsilon_l^\text{max}\equiv\text{max}(|V_l(z_0)|,|V_l(z_1)|)$, which is explained in the following.

For energies $\epsilon_{nl}>\epsilon_l^\text{max}$, a solution $\chi_{nl}$ of 
	Eq.~(\ref{laplaceeqq}) is fully extended across the TINC and extends into both leads, and $n$ is a continuous index; in the limit of zero slope, this index is simply $n=k_z$.
For $\epsilon_l^\text{min}\leq\epsilon_{nl}\leq \epsilon_l^\text{max}$, 
	any solution $\chi_{nl}$ extends into one of the leads, while decaying exponentially on the other side, 
	and again $n$ is continuous.
For intermediate/high magnetic fields potential wedges are present, see, for instance, the last panel in Fig.~\ref{fig:nanocone}, and $\epsilon_l^\text{min}$ represents the depth of such a wedge. Thus, the possibility for bound states within the effective potential with energies $\epsilon_{nl}<\epsilon_l^\text{min}$ arises.  
As pointed out in Ref. \cite{Kozlovsky2020}, the energies of these bound states are given by the Dirac LL energies
\begin{equation}
E_n=\text{sgn}(n)\frac{\hbar v_F}{l_B}\sqrt{2|n|},\hspace{1cm}n\in\mathbb{Z}\label{LLens},
\end{equation}
where $l_B\equiv\sqrt{\hbar/(e|B_\perp|)}$ is the magnetic length and $B_\perp$ is the magnetic field component perpendicular to the surface of the TINC, cf.~Fig.~\ref{fig_cones}(a). This outcome is corroborated in the following with numerical results and an intuitive physical picture.

\begin{figure}
	\centering
	\includegraphics[width=\columnwidth]{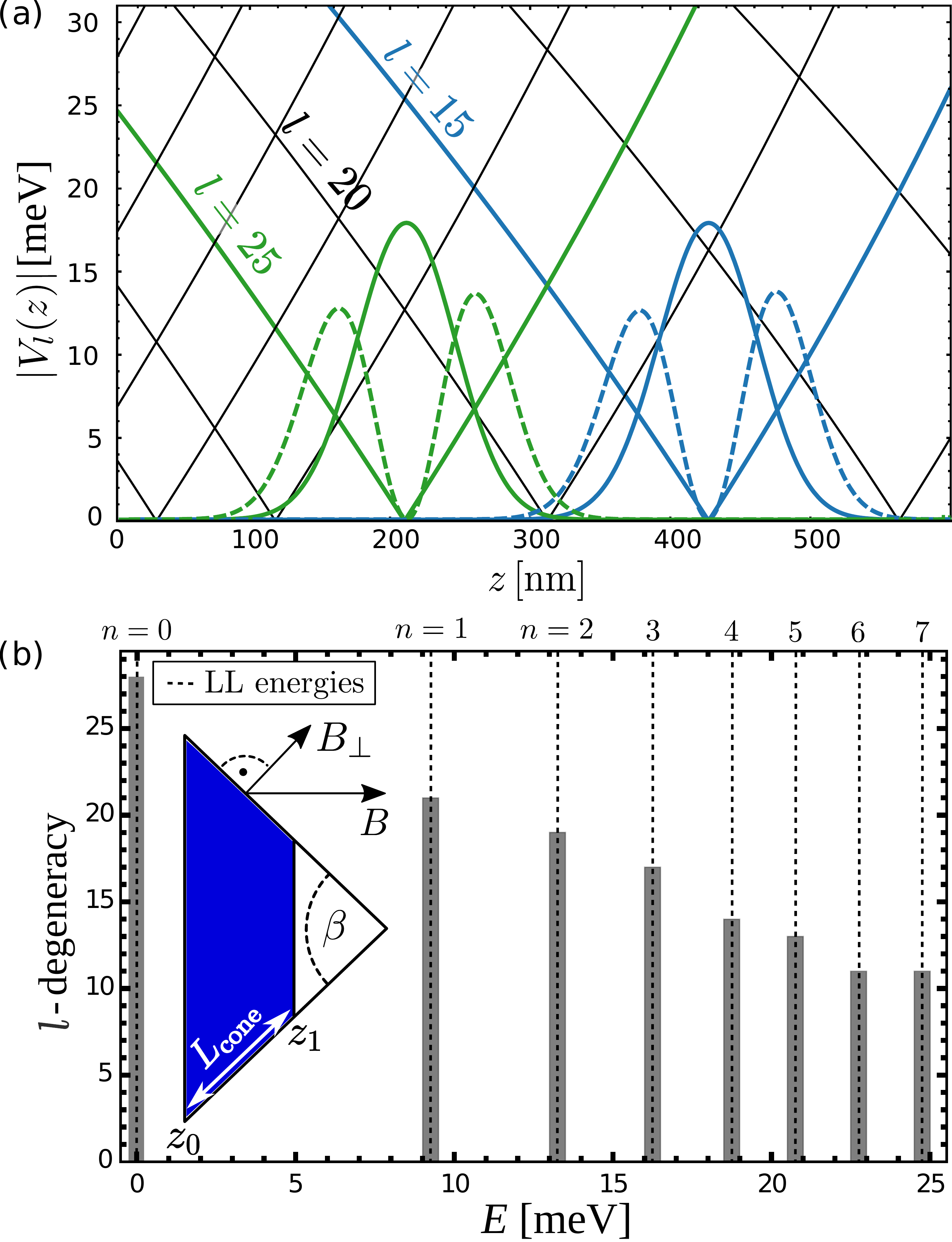}
	\caption{(a) Effective potentials $|V_l|$ for cone parameters $z_0=0$, $z_1=594.7\,$nm, $R_0=2R_1=156.6\,$nm, and $B=2\,$T, which yields $l_B\approx 50$~nm, $L_{\rm cone}=600$~nm. Note that $|V_l|$ are plotted for every fifth $l$-value only.  Additionally, the eigenstate probability distributions [for the solutions of Eq.~\eqref{Direq}] $|\chi_{n=0,l=15}(z)|^2$ (blue solid line), $|\chi_{n=1,l=15}(z)|^2$ (blue dashed line), $|\chi_{n=0,l=25}(z)|^2$ (green solid line) and $|\chi_{n=1,l=25}(z)|^2$ (green dashed line) are shown (using arbitrary units). The bound states are identified as QH states (see main text). 
	(b) Bar plot for the bound state energies $\epsilon_{nl}$. The width of the bars is $0.4\,$meV. Dashed vertical lines give the analytical values of the Dirac LL energies with the perpendicular magnetic field component $|B_\perp|=B\sin(\beta/2)$.
	The inset shows a schematic side view of the TINC, with opening angle $\beta$.
	}
	\label{fig:QH_states}
\end{figure}

In Fig.~\ref{fig:QH_states}, we choose parameters $\{R_0,R_1,z_0,z_1,B\}$ with 
$l_B \ll L_{\rm cone} \equiv (z_1-z_0)/\cos(\beta/2)$ ($l_B\approx 50$~nm, $L_{\rm cone}=600$~nm), which highlights the formation of LLs. Here, $L_{\rm cone}$ is the arc length along the TINC and $\beta$ the opening angle.
Figure~\ref{fig:QH_states}(a) shows the effective potentials $|V_l|$ for the corresponding TINC together with the probability distributions of the eigenstates $|\chi_{nl}(z)|^2$ for the two wedges $l=15$ and $l=25$, where the solid (dashed) line corresponds to the $n=0$ ($n=1$) state. Note that, for clarity of the Figure, $|V_l|$ is not plotted for all $l$-values (unlike in Fig. \ref{fig:nanocone}).
Figure \ref{fig:QH_states}(b) shows a bar plot which counts eigenenergies $\epsilon_{nl}$ in an energy window of $0.4\,$meV. Here, all energies $\epsilon_{nl}$ were used for which the states $\chi_{nl}$ are bound states \textit{on} the TINC, \ie reside between $z_0$ and $z_1$.
As expected, we observe a large degeneracy in $l$ at the Dirac LL energies $E_n$ (up to numerical precision) marked with vertical dashed lines. 
Moreover, the eigenstate probability distributions $|\chi_{nl}(z)|^2$ show one maximum for $n=0$ and two maxima for $n=1$, as expected from quantum Hall (QH) states which derive from a harmonic oscillator equation (for which a state with index $n$ has $n+1$ maxima).
Hence, we can indeed identify the bound states of the effective potential wedges $|V_l|$ as QH states, and all QH states (labeled by $n,l$) for a given $n$ together form the $n$-th LL. 

The intuitive physical picture is the following: The perpendicular magnetic field component is constant throughout the cone, which means that the 2D Dirac electrons on the surface are subject to a homogeneous magnetic field and thus form LLs. The only condition which needs to be fulfilled is that the magnetic length is small compared to the length of the cone, such that the QH states (in classical terms cyclotron orbits) fit onto the TINC. This is equivalent to the condition that effective potential wedges form within the TINC. 
These arguments are also reflected in the degeneracies of the LLs in Fig.~\ref{fig:QH_states}(b), which is given by the height of the bars. Quantum Hall states with larger $n$ extend more in space, and thus less QH states fit onto the cone. Consequently, the degeneracy decreases with increasing $n$.

\subsection{Transport through a TI nanocone}
\label{TINCtrans}

The setup we consider for transport is a TINC connected to cylindrical, highly-doped TI leads in a coaxial magnetic field, see Fig.~\ref{fig_cones}(a).
Its transport properties are determined by the states available at a given energy, 
\ie those discussed in Sec.~\ref{TINCstruc}:
\begin{enumerate}[label={[\Alph*]}]
	\item At high energies ($\epsilon_{nl}>\epsilon_l^\text{max}, n$ continuous), states are fully extended across the TINC and hybridize with both leads.
	\item At intermediate energies ($\epsilon_l^\text{min}\leq\epsilon_{nl}\leq \epsilon_l^\text{max}, n$ continuous), states couple strongly to one of the leads and weakly if at all to the other.
	\item If $\epsilon_{nl}<\epsilon_l^\text{min}$, quasi-bound states centered at $\tilde{z}_l(B)$ exist.  
	For $\epsilon_{nl}\ll\epsilon_l^\text{min}$ their energy coincides with LL states $\epsilon_{nl}=E_n$.
	Closer to the potential threshold, $\epsilon_{nl}\lesssim\epsilon_l^\text{min}$, the tail of the wave function enters the leads, and $\epsilon_{nl}\lesssim E_n$.
\end{enumerate}
The considerations above, together with knowledge from Sec.~\ref{subsec_physical_picture}, allow us to make qualitative predictions for the conductance $G$ as a function of the lead Fermi level $E_F$ and the coaxial magnetic field $B$. These predictions will be confirmed by numerical transport simulations later on.

\textit{Low magnetic field --}
Inspecting the effective potential in Fig.~\ref{fig:nanocone}, one expects $G(E_F)$ to be characterized by steps centered around energies $\epsilon_l^\text{max}$, since the effective potential of the left (right) lead is given by $|V_l(z_0)|$ ($|V_l(z_1)|$).

\textit{Intermediate/high magnetic field --}
In the situation shown in Fig.~\ref{fig:nanocone} for $B=0.5\,$T, only the potential wedges belonging to $l=7,8$ feature thresholds above the first LL. Consequently, the first LL can only form in the central part of the TINC, and higher LLs are absent.
Thus we choose the TINC parameters from Fig.~\ref{fig:QH_states}, where LLs consisting of many QH states form and their role in transport is enhanced.
\commentout{
\red{To a very good approximation, the center of any quasi-bound state forming in a given potential wedge is located at $\tilde{z}_l$. Therefore, intuitively speaking, a LL in a wedge $|V_l|$ can form if the corresponding magnetic length easily fits onto the TINC, \textit{i.e.}, if $\text{min}(|\tilde{z}_l(B)-z_1|,|\tilde{z}_l(B)-z_0|)\gg l_B$.
In the language of the effective potential, one needs to (i) keep the potential wells deep enough (large $\epsilon_l^\text{min}$) while (ii) keeping the LL spacing low enough, such that LLs with $n>0$ can form across as large a fraction of the TINC's surface as possible. The first condition is achieved for small radii and large $B$, cf. Eq. (\ref{effpot}), while the second condition is assisted by low $B$ and low $\mathcal{S}$. From this intricate interplay, we find that TINCs realized in relatively narrow wires ($R\approx100\,$nm) and of small slope ($\mathcal{S}\sim1/10$) at not too high magnetic fields ($B\approx2\,$T) are ideal candidates.}
}
\begin{figure}
	\centering
	\includegraphics[width=\columnwidth]{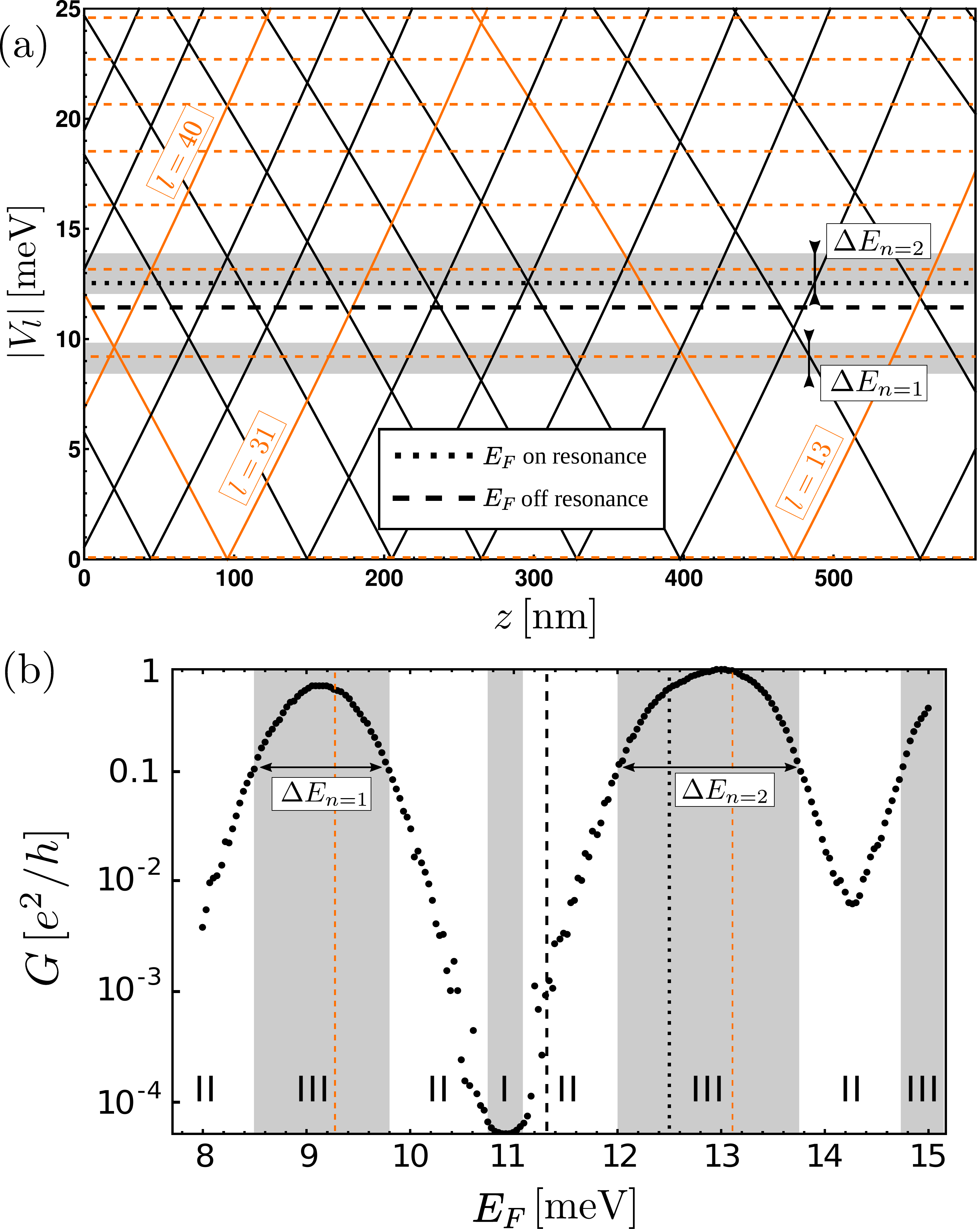}
	\caption{Transport through a TINC in intermediate/high magnetic field for parameters as in Fig.~\ref{fig:QH_states}. (a) Effective potentials $|V_l|$ on the TINC. Note that $|V_l|$ are plotted for every third $l$-value only. LL energies are indicated by orange dashed lines. The shaded stripes denote typical broadening of the LLs, as extracted from the numerical results in panel (b). (b) Logarithmic plot of the disorder-averaged conductance around the first and second LL, as a function of lead Fermi level $E_F$ and calculated using \textsc{Kwant}. Vertical lines have the same meaning as in panel (a). The labels (I, II, III) serve to explain the transport regimes of the TI dumbbell (see Sec.~\ref{dumbbell} and Fig. \ref{fig:dumbbell}).}
	\label{fig:coneLLs}
\end{figure}
The effects on transport of such strong LL quantization are presented in Fig.~\ref{fig:coneLLs},
showing $|V_l|$ and $G(E_F)$ for a TINC geomentry defined in the caption and a coaxial magnetic field of $B=2\,$T. 
Here disorder is added, which couples different $l$-modes (\ie QH states).  This causes a broadening 
of the LLs, sketched in Fig.~\ref{fig:coneLLs}(a) as gray shaded areas around their central (ideal) energies 
(horizontal orange lines).

Independently of the value of $E_F$, a lead electron can enter the outer TINC region via states of type [B].
However, $E_F$ determines whether it can enter the central region:
\begin{itemize}
	\item Off-resonance -- If $E_F$ is far away from a LL energy, an electron in the outer TINC region (where no LL forms), cannot find states for elastic transport further into the TINC, hence the conductance is suppressed. 
	\item On-resonance -- If $E_F$ lies within a disorder-broadened LL, characterized by a certain width $\Delta E_n$, an electron in the outer TINC region can, via disorder-induced scattering, be transferred to a state of type [C]. From there, it can travel elastically through the central TINC region (via disorder-coupled QH states), such that the conductance is finite. 
\end{itemize}
For visualization, consider a lead electron from the left at $E_F<E_2-\Delta E_2/2$ [black dashed line in Fig. \ref{fig:coneLLs}(a)]. It can enter the TINC via states of type [B], with $31\lesssim l\lesssim40$. Then, however, elastic transport is obstructed, because potential wedges in the center of the TINC only host states of type [C], with $\epsilon_{nl}\approx E_n$. In contrast, if $E_2-\Delta E_2/2\lesssim E_F\lesssim E_2+\Delta E_2/2$ [black dotted line in Fig. \ref{fig:coneLLs}(a)], a lead electron from the left, after accessing the TINC via modes $31\lesssim l\lesssim 40$, can elastically tunnel through the core region of the TINC, via states of type [C] which exist for $13\lesssim l\lesssim 30$, and exit the TINC on the other side.

We conclude that, at low energies, the TINC is \textit{transparent} for $E_F\approx E_n$, while it is \textit{opaque} for $E_F$ in between two consecutive LLs. 
Fig.~\ref{fig:coneLLs}(b) shows the TINC conductance around the first and second LLs, calculated using the \textsc{kwant} \cite{groth2014} software. The resonant conductance peaks, already numerically obtained in Ref.~[\onlinecite{Kozlovsky2020}], are explained in an intuitive way by the microscopic picture outlined above. Highly-doped leads were used for the calculations, 
together with Gaussian-correlated disorder, \smash{$\braket{V(\mathbf{r})V(\mathbf{r}')}=K\hbar v_Fe^{-|\mathbf{r}-\mathbf{r}'|^2/2\xi^2}/(2\pi\xi^2)$}, 
with the (dimensionless) disorder strength $K$ and the correlation length $\xi$.
For numerical results presented throughout this paper, averages were taken over 600 disorder configurations with $K=0.1$. 
(For more details on the methodology of the numerics, see Ref.~[\onlinecite{Kozlovsky2020}] or  Appendix~\ref{AppConductanceKwant}).
Importantly, Fig. \ref{fig:coneLLs}(b) is also the starting point for describing transport in a TI dumbbell.

\begin{figure}
	\centering
	\includegraphics[width=\columnwidth]{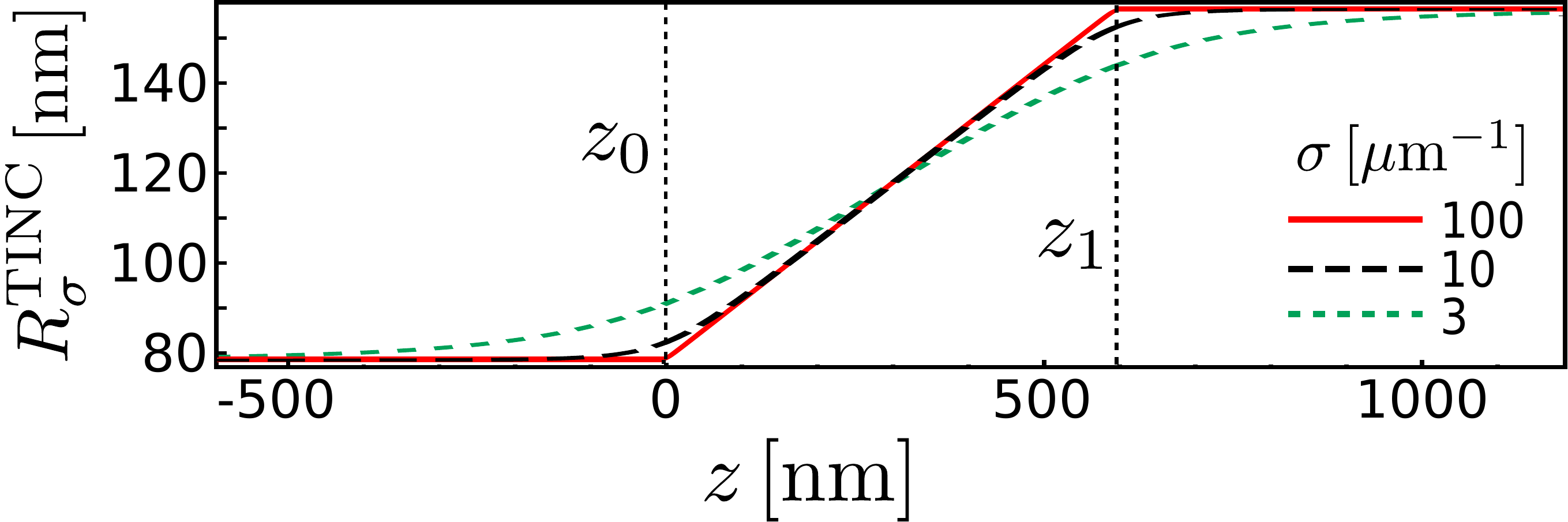}
	\caption{Radius of a smoothed TINC as given by Eq.~(\ref{coneapprox}) for three values of $\sigma$, with parameters chosen such that the (perfectly sharp) TINC geometry used in Figs. \ref{fig:QH_states} \& \ref{fig:coneLLs} is recovered in the limit $\sigma\to\infty$.}
	\label{fig:radius}
\end{figure}

\subsection{Smoothed TI nanocone}
\label{sec:smoothed_TINC}

Consider now a more realistically shaped TINC with smooth connections to the leads, cf. Fig.~\ref{fig_cones}(c).
The parametrization of the corresponding radius $R(z)$, cf. Fig.~\ref{fig:radius}, is given by Eq.~\eqref{coneapprox}. 
The smoothing strength is determined by the parameter $\sigma$, where a small (large) $\sigma$ corresponds to strong (weak) smoothing.

\begin{figure}
	\centering
	\includegraphics[width=\columnwidth]{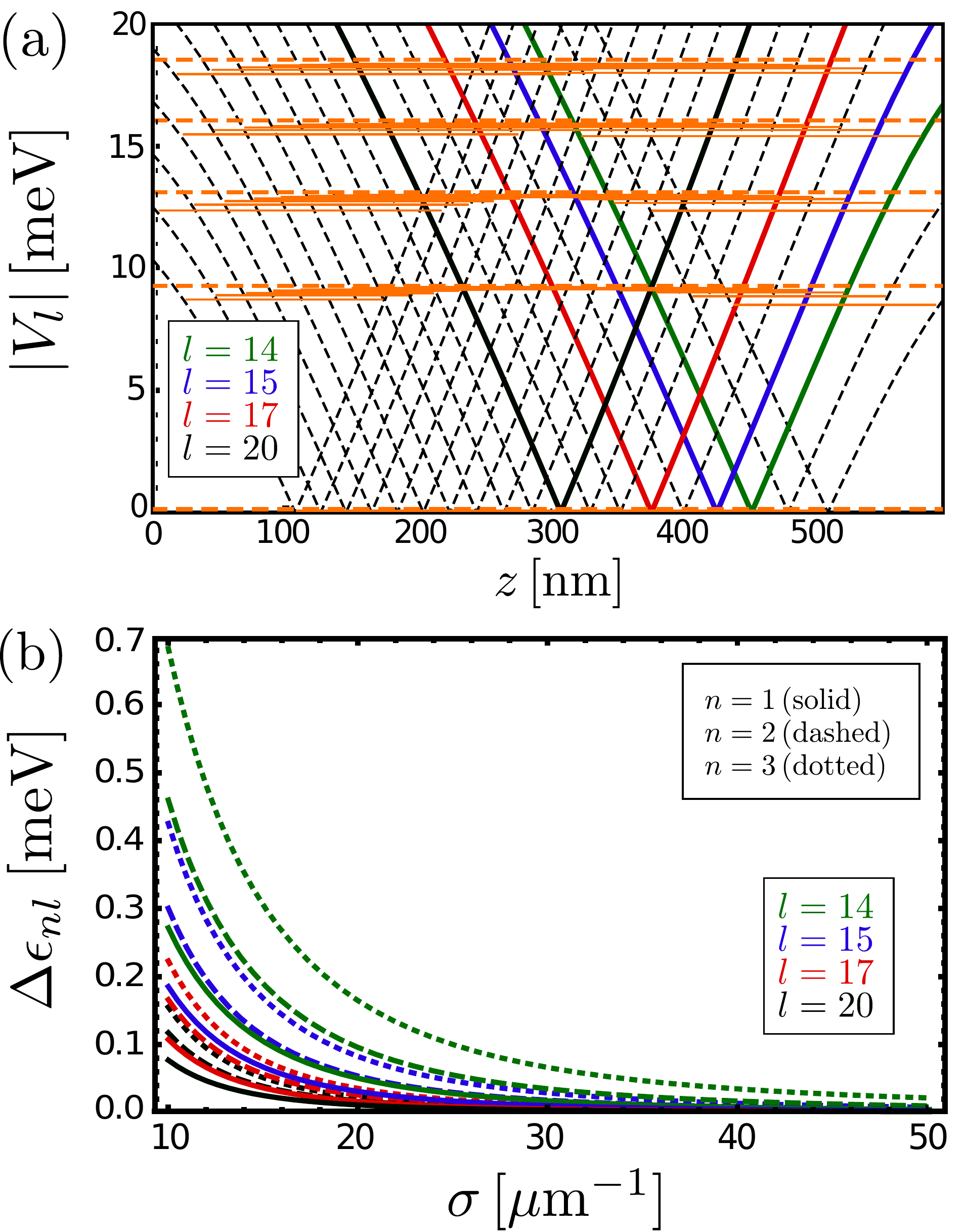}
	\caption{(a) Effective potentials for a smoothed TINC, with the same parameters as in Fig. \ref{fig:coneLLs}, but lowering the value of $\sigma$ in Eq. (\ref{coneapprox}) to $\sigma=10\,\mu$m$^{-1}$. Orange dashed lines represent LL energies $E_n$ in the limit $\sigma\to\infty$ of a perfect TINC, while solid orange lines are bound state energies $\epsilon_{nl}$. (c) Difference in energy $\Delta\epsilon_{nl}(\sigma)$ between LLs of the perfect TINC [dashed orange lines in panel (a)] and bound state energies $\epsilon_{nl}$ of the potential well $|V_l|$ [solid orange lines in panel (a)].}
	\label{fig:smoothcone}
\end{figure}

In the low magnetic field regime there is no qualitative difference with the ideal TINC,
and conductance steps centered around energies $\epsilon_l^\text{max}$ are expected. In the remainder of the present section, we focus on intermediate and high fields, starting as usual by solving Eq.~(\ref{laplaceeqq}).
In Fig. \ref{fig:smoothcone}, we show the effect of smoothing for the same system parameters as in Fig.~\ref{fig:coneLLs}, meaning that the TINC from Fig.~\ref{fig:coneLLs} is recovered in the limit $\sigma\rightarrow\infty$.
When the TINC is smoothed (by decreasing $\sigma$), the value of $|B_\perp|$ gets lowered, the effect being considerably stronger in the vicinity of the leads than in the middle of the TINC, cf.~Fig.~\ref{fig:radius}.

In the language of the effective potential, this means that a given wedge $|V_l|$ shifts and gets distorted (mostly its lead-facing branch gets lowered), see Fig.~\ref{fig:smoothcone}(a). This effect is stronger for wedges close to the leads. Therefore, the smoothing can have two different effects on a given LL bound state.
(i) The bound state disappears. This is relevant for states close to the leads.
(ii) The bound state survives but gets lowered in energy because of reduced $B_\perp$ (increasing magnetic length). 

Thus, upon smoothing, the $l$-degeneracy of the LLs, present for the perfect TINC in Sec.~\ref{TINC}, is lifted, see Fig.~\ref{fig:smoothcone}(a). Note that the zeroth LL stays $l$-degenerate, since it is not affected by the smoothing. For all states belonging to class (ii), we define the decrease in energy $\Delta\epsilon_{nl}(\sigma)\equiv \epsilon_{nl}(\sigma\to\infty)-\epsilon_{nl}(\sigma)>0$, where $\epsilon_{nl}(\sigma\to\infty)=E_n$. It is plotted for LL indices $n=1$ to $n=3$ in Fig. \ref{fig:smoothcone}(b), for a selected number of angular momenta.

Consider now the transport characteristics of the smoothed TINC.
An electron from the lead at given $E_F$ can typically proceed a bit further into the TINC than in the limit $\sigma\to\infty$. This is due to the lowering of the lead-facing effective potential branches, see Fig. \ref{fig:smoothcone}(a).
However, the energies of states in the TINC center stay practically the same as in the limit $\sigma\to\infty$. Thus, transport through the TINC is still suppressed for $E_F$ placed in between LLs.

Pursuing this line of thought, we can predict another interesting transport regime in smoothed TINCs. If the decrease in energy of bound states relatively close to the leads [for example the state $(n,l)=(2,14)$ in Fig. \ref{fig:smoothcone}(a)] becomes large enough ($\Delta\epsilon_{nl}(\sigma)>\Delta E_n/2$), these energies exit the disorder-broadened transport channel.  
Such states are no longer available for elastic transport, even though the Fermi level is tuned "on resonance'',
\ie $E_F\approx E_n$. This phenomenon can be achieved in our setup using relatively strong magnetic fields $B\gtrsim5\,$T. Consequently, magnetic barriers arise close to the TINC ends, such that \textit{a single} (smoothed) TINC becomes a quantum dot-like object. If a gate electrode is attached to the TINC, this may lead to Coulomb blockade-type physics \cite{Kozlovsky2020}.

In summary, the qualitative form of the conductance $G(E_F)$ shown in Fig. \ref{fig:coneLLs}(b) is \textit{robust} against variations of the geometry, as long as the local QH state energies lie within the disorder broadening. For a smoothing [as defined by Eq.~\eqref{coneapprox}] strength for which the QH states close to the leads are moved beyond the disorder broadening, magnetic barriers appear and we expect transport to be dominated by Coulomb blockade-like physics, as discussed in Ref. \cite{Kozlovsky2020}.


\section{Magnetotransport characteristics of a TI dumbbell}
\label{dumbbell}

As an example of shaped TINWs beyond the relatively simple TINC, we now consider a TI dumbbell [see Fig.~\ref{fig_cones}(b)], which is representative for a mesoscopic TI nanowire constriction and hosts a rich variety of magnetotransport regimes.
For simplicity we take the TI dumbbell to be composed of two symmetrically arranged TINCs ($R_2=R_1$, $R_3=R_0$), 
cf.~Fig.~\ref{fig_cones}(b), each with the same parameters as used in Figs.~\ref{fig:QH_states} and \ref{fig:coneLLs}.
The length of the intermediate cylindrical part is chosen as $L\equiv z_2-z_1=(z_1-z_0)/2$.

\subsection{Low magnetic field}

\begin{figure}
	\centering
	\includegraphics[width=\columnwidth]{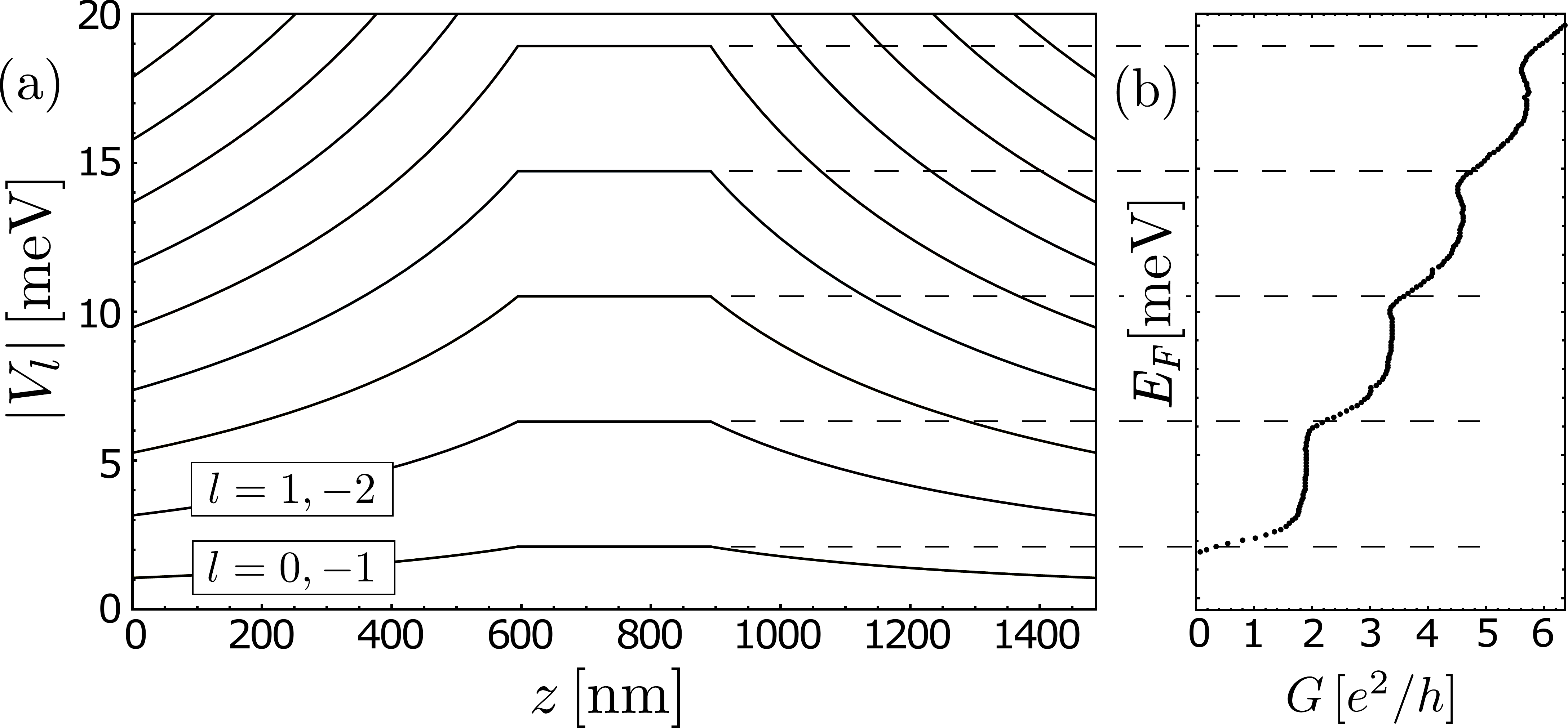}
	\caption{Transport through a TI dumbbell for $B=0$. Here we choose $z_0=0$, $z_1=594.7\,$nm, $z_2=3z_1/2$, $z_3=5z_1/2$, and $R_0=R_3=2R_1=2R_2=156.6\,$nm. (a) Effective potentials $|V_l|$ and (b) disorder-averaged conductance as a function of the lead Fermi level $E_F$, calculated using \textsc{kwant}.}
	\label{fig:DBcond}
\end{figure}

The effective potentials feature an $l$-degeneracy for $B=0$ [see Fig.~\ref{fig:DBcond}(a) and recall Fig.~\ref{fig:nanocone}], leading to a conductance profile as shown in Fig. \ref{fig:DBcond}(b). This $l$-degeneracy gets lifted for $B\neq0$. When this happens, the precise form of the conductance $G(E_F)$ depends on the particular value of $B$, but its qualitative structure (smoothed steps originating from mode opening) stays the same as long as $B_\perp$ is too weak for LLs to form on the two TINCs. (Although the dumbbell considered there is of slightly different dimensions than the one discussed in the present Section, this degeneracy lifting, as well as the qualitatively unchanged conductance profile, can be observed explicitly in Fig. \ref{G_smoothedDB}, which is plotted in the low $B$ regime.) Due to the mirror symmetry with respect to the plane $z=(z_3-z_0)/2$, these steps are located at the same values $E_F=\epsilon_l^\text{max}$ that one would expect for a single TINC.  

\subsection{Intermediate/high magnetic field}

In this regime, a simple tuning of the magnitude of $B$ allows access to three scenarios: (I) current suppression, (II) Coulomb blockade, and (III) resonant transmission.

\begin{figure}
	\centering
	\includegraphics[width=\columnwidth]{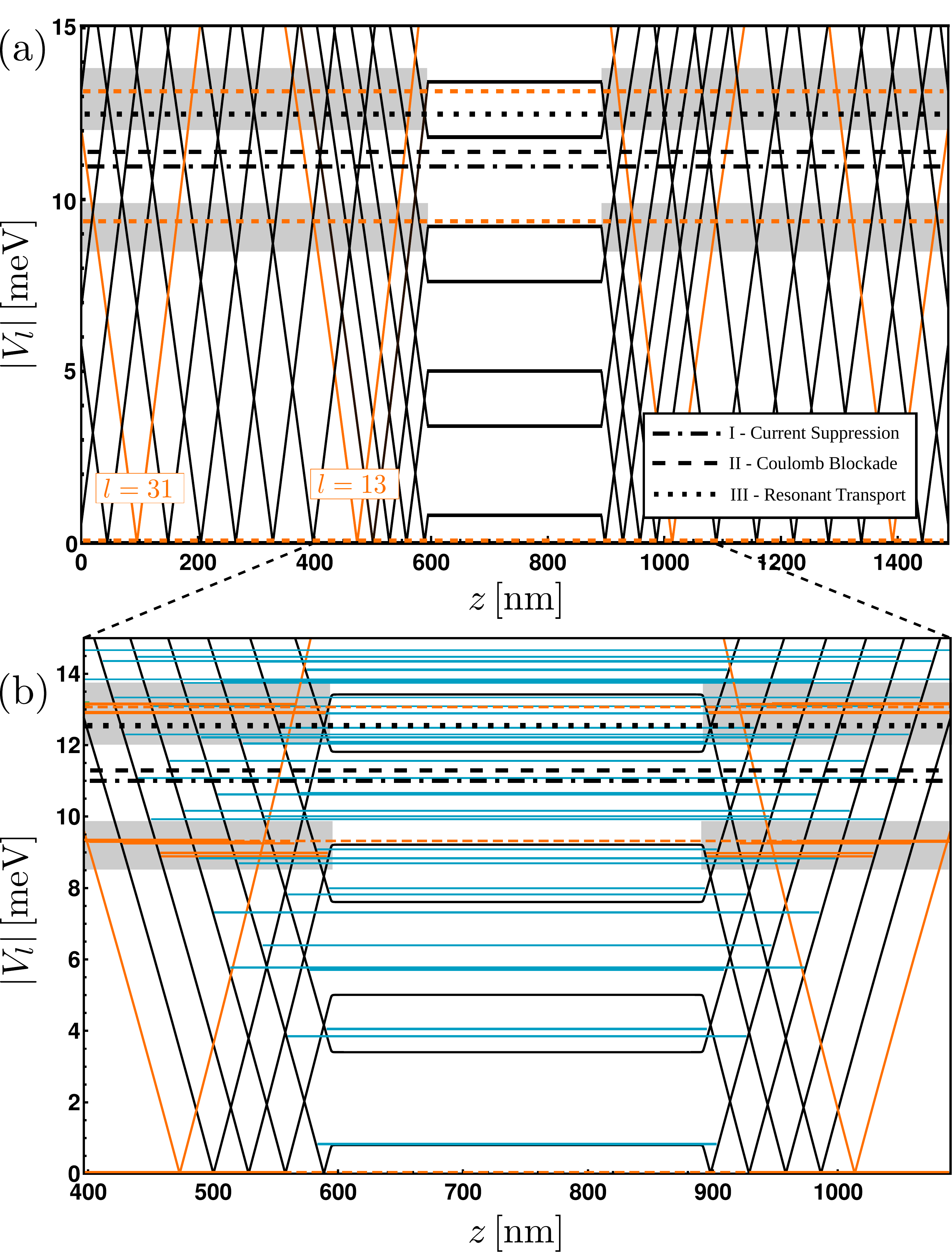}
	\caption{(a) Effective potentials for a TI dumbbell for $B=2\,$T and parameters as in Fig.~\ref{fig:DBcond}. Note that $|V_l|$ are plotted for every third $l$-value only, except in the important central region, which is the focus of this Figure. (b) Zoom into the central region. Energies of eigenstates residing on the central cylinder (on the TINCs) 
	are marked by blue (orange) lines.}
	\label{fig:dumbbell}
\end{figure}

Given our choice of parameters, the effective potential landscape on the dumbbell's left side is the one shown in Fig.~\ref{fig:coneLLs}(a), followed by a constant $|V_l|$ in the cylindrical center and the mirrored version of Fig.~\ref{fig:coneLLs}(a) on the right, see Fig.~\ref{fig:dumbbell}(a).
Potential wells appear in the central region,
which may host bound states whose coupling to the leads depends on the transparency of the TINCs. This transparency is the same for both TINCs,  and governed by the conductance $G(E_F)$ shown in Fig.~\ref{fig:coneLLs}(b), because the ladder of LLs depends only on the absolute value of $B_\perp$.

Figure \ref{fig:dumbbell}(b) depicts the energies [obtained from solving Eq.~\eqref{laplaceeqq} in the presence of cylindrical TI leads] corresponding to QH states on the TINCs as orange lines, and those corresponding to states confined between the two TINCs as blue lines. As a guide to the eye, the extent of the lines in the $z$-direction is chosen such that their ends touch the potential well they belong to (this encodes the $l$-value of a given energy level). Within each given $l$-potential well, the level spacing 
(distance between blue lines) can be controlled by the longitudinal confinement and is proportional to $1/L$.
However, it is in general not possible to define a constant level spacing 
characterizing the whole central region, as states belonging to different $l$-potentials may cluster.

Using Fig.~\ref{fig:coneLLs}(b), three different transport regimes can be identified, 
depending on the lead Fermi level position relative to the LL energies.
\begin{enumerate}[label={(\Roman*)}]
	\item If $E_F$ is placed such that $G\approx0$ ($G$ being the conductance through a single TINC), both TINCs act as strong barriers, and transport through the dumbbell is suppressed. This is indicated by the dashed-dotted line in Fig. \ref{fig:dumbbell}.
	\item If $E_F$ is placed such that the TINC conductances are $0<G\ll e^2/h$, the central cylindrical region can be viewed as a quantum dot weakly coupled to the leads.  As we discuss below, this leads to conductance oscillations of Coulomb
	blockade type once the central region is gated. This is indicated by the dashed line in Fig. \ref{fig:dumbbell}.
	\item
	 Both TINCs are transparent ($G\sim e^2/h$) for a lead Fermi level fulfilling $E_F\approx E_n$. 
	 If $E_F$ is in addition in resonance with a (disorder broadened) bound state of the central cylinder (blue line), we expect a finite conductance. Otherwise the transmission is suppressed. This is indicated by the dotted line in Fig. \ref{fig:dumbbell}.
\end{enumerate}

In the remainder of this Section, we focus on case (II). 

\subsubsection*{Coulomb blockade in the TI dumbbell}

If we assume a gate electrode to be applied to the central cylindrical region, a decisive quantitiy is the charging energy $E_C=e^2/C$
, where the capacitance $C$ depends on the experimental setup (the geometry and the materials, including the dielectrics). Some typical values for $C$ are provided by the literature.
For experiments on strained HgTe TINWs \cite{ziegler2018} of dimensions comparable to our situation, a numerical solution of the Poisson equation in the presence of \ce{SiO2}/\ce{Al2O3} dielectrics and a gold top gate yields an effective capacitance per surface area $C_\text{eff} \approx 4\cdot10^{-4}\,\text{F}\,\text{m}^{-2}$. A different experiment, studying TI quantum dots based on \ce{Bi2Se3} TINWs on a \ce{SiO2}/Si substrate \cite{cho2012}, found $C=2\cdot10^{-17}\,$F for a surface area of $8.6\cdot10^{-14}\,$m$^2$, corresponding to $C_\text{eff}=2.3\cdot10^{-4}\,\text{F}\,\text{m}^{-2}$. We can thus estimate the charging energy for the TI dumbbell by $E_C=e^2/(2\pi R_1LC_\text{eff})$. 
For our example, we choose a value of $E_C=1.5\,$meV, which corresponds to $C_\text{eff}=7.3\cdot10^{-4}\,\text{F}\,\text{m}^{-2}$, the same order of magnitude as in the above experiments. This choice results in $E_C \approx \delta \epsilon$; here, $\delta \epsilon$ is the typical level spacing in the central region, defined as follows: Each $\epsilon_{nl}$ quasi-bound state in the central region [blue lines in Fig.~\ref{fig:dumbbell}(b); $n$ counts the quasi-bound states within a given well $|V_l|$] has a broadening $\Gamma_{nl}$ determined by its coupling to the leads.
Blue lines clustering such that their energy distance satisfies
\begin{equation}
|\epsilon_{nl}-\epsilon_{n'l'}|< \frac{1}{2}{\rm max}\left\{\Gamma_{nl},\Gamma_{n'l'}\right\}
\end{equation}
cannot be resolved, yielding effectively a single central region level (which can be multiply occupied).
The spacing $\delta\epsilon$ is taken among such central region levels, be they single $\epsilon_{nl}$ levels
or clusters as just defined, and should be viewed as an order-of-magnitude estimate 
obtained by inspecting Fig.~\ref{fig:dumbbell}(b).

In the presence of a gate electrode, we expect Coulomb blockade oscillations \cite{beenakker1991} in the current-gate voltage characteristics. The particular Coulomb blockade regime is determined by the ratio of the three energy scales $\delta\epsilon$, $k_BT$ and $\Gamma_{nl}$. An order of magnitude estimate for $\Gamma_{nl} = \hbar / \tau_{nl}$
follows from the dwell time $\tau_{nl}$ in the central region.  
The two side TINCs can be treated as ``black boxes'' with a transmission given by the conductance from Fig.~\ref{fig:coneLLs}(b), such that the system is mapped to a quasi-1D double-barrier structure of inner length $L$.
Estimating $\tau_{nl}$ requires two ingredients: 
(i) The average distance (in the $z$-direction) covered by an $\epsilon_{nl}$ quasi-bound state 
is $d=L\frac{1+{\cal R}_2}{1-{\cal R}_1 {\cal R}_2}$, with ${\cal R}_i, i=1,2$ the reflection probabilities at 
the individual barriers.
In our symmetric setup we have ${\cal R}_1={\cal R}_2\equiv{\cal R}$, and $d=L/{\cal T}$, with ${\cal T}=1-{\cal R}\ll1$.
(ii) The average velocity (in the $z$-direction) can be estimated as 
\begin{equation}
\hbar v_{nl}\sim\partial\epsilon_l(k_{z,n})/\partial k_{z,n}
\end{equation}
where
\begin{equation}
\epsilon_l(k_{z,n})=\hbar v_F[(l+1/2-\Phi/\Phi_0)^2/R^2+k_{z,n}^2]^{1/2}
\end{equation}
is the band structure of a cylindrical TINW, with $k_{z,n}=(\pi/L)n$ the wavevector values
corresponding to a given quasi-bound state $n$.
It follows 
\begin{equation}
\Gamma_{nl}=\frac{\hbar}{\tau_{nl}}\sim \frac{\hbar v_{nl}}{d} 
\sim
\frac{\cal T}{L}\frac{(\hbar v_F)^2 k_{z,n}}{\epsilon_l(k_{z,n})}.
\end{equation}
Thus, using a typical value ${\cal T}=5\cdot10^{-3}$ [cf.~Fig.~\ref{fig:coneLLs}(b)], one has $\Gamma_{nl}\sim 5\,\mu$eV.
While this implies that a few of the $\epsilon_{nl}$ levels from Fig.~\ref{fig:dumbbell}(b) form clusters,
spacing among the latter is such that at temperatures around $T\approx 0.5\,$K various central region levels
(single $\epsilon_{nl}$ or clusters thereof) are resolved, 
\ie the condition $\Gamma_{nl}\ll k_BT\ll\delta\epsilon$ is fulfilled.

\begin{figure}
	\centering
	\includegraphics[width=\columnwidth]{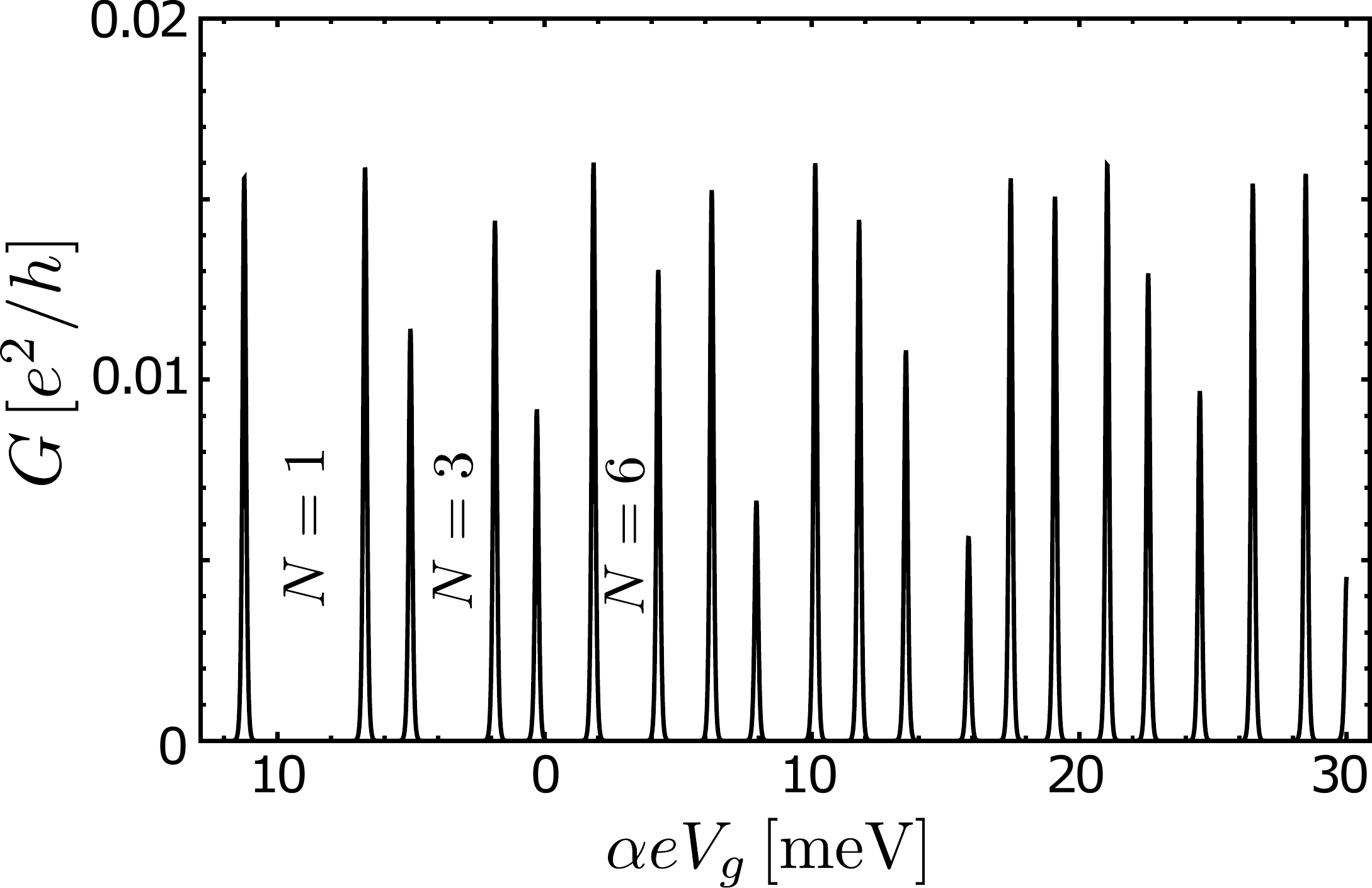}
	\caption{Coulomb blockade oscillations for a (surface) quantum dot formed due to magnetic confinement in a TI dumbbell with the same parameters as in Fig. \ref{fig:dumbbell}. The conductance is calculated using Eq. (\ref{conduct}), at a temperature $T=0.5\,$K, lead Fermi level $E_F=11.3\,$meV and charging energy $E_C=1.5\,$meV.}
	\label{fig:CBosc}
\end{figure}

In this regime, only a single level contributes to transport significantly, and the conductance is given by \cite{beenakker1991}
\begin{equation}
G=\frac{e^2}{h}\Gamma_{nl}\frac{f(\Delta_{nl})[1-f(\Delta_{nl})]}{2k_BT}.
\label{conduct}
\end{equation}
The quantity $\Delta_{nl}\equiv\epsilon_{nl}-E_F+(N_{nl}-1/2)E_C-\alpha eV_g$ entering the Fermi-Dirac distribution $f(x)=1/(1+e^{x/k_BT})$ leads to a conductance peak whenever $\Delta_{nl}=0$. Here $N_{nl}$ is the number of electrons on the quantum dot, \ie the number of levels with $\epsilon_{n'l'}<\epsilon_{nl}$; The proportionality constant $\alpha$ between gate voltage $V_g$ and the associated electrostatic energy is, like $C$, a function of the capacitance matrix \cite{beenakker1991} and needs to be determined experimentally \cite{Ihn}. 
The resulting Coulomb blockade oscillations for the dumbbell from Fig.~\ref{fig:dumbbell} are shown in Fig.~\ref{fig:CBosc}, taking $E_F=11.3\,$meV [dashed black line in Figs.~\ref{fig:coneLLs} and \ref{fig:dumbbell}],  
$E_C=1.5\,$meV, $T=0.5\,$K and ${\cal T}=5\cdot10^{-3}$. The fluctuations of conductance peak positions reflect the level spacings of the confined states living in the central cylindrical region.

Note that our discussion of Coulomb blockade, and more generally of  all the transport regimes considered above, aims at identifying \emph{intrinsic} geometry-induced features.  Therefore, the role of  additional system-specific properties, \eg voltage ripples \cite{Munoz_2016} that might affect the Coulomb blockade oscillations, are not considered in our theoretical model (\ref{surfdir}).
	
\subsection{Smoothed TI dumbbell}
\label{smooth}


The knowledge about a single smoothed TINC, discussed in Sec.~\ref{sec:smoothed_TINC}, is straightforwardly generalized to more complex shaped TINWs. In particular, the smoothed TI dumbbell from Fig.~\ref{fig_cones}(c) 
(plotted for $\sigma=15\,\mu$m$^{-1}$) still exhibits qualitatively the same transport behavior as the model junction discussed in the course of the present Section. 
As for the smoothed TINC, a new transport regime emerges at rather high magnetic fields. In this case, the smoothed dumbbell is still a double-barrier structure off resonance. However, it is a quadruple-barrier structure on resonance, \ie~a triple quantum dot, since each TINC features a pair of potential barriers.

\begin{table*}[ht]
	\centering
	\begin{tabular}{P{0.15\linewidth}||P{0.15\linewidth}P{0.15\linewidth}|P{0.15\linewidth}P{0.17\linewidth}P{0.17\linewidth}}
		& $\mathbf{B}\perp\hat{z}$ &  & $\mathbf{B}\parallel\hat{z}$ &  & \\
		\hline
		& weak $B$  & strong $B$ & weak $B$	& intermediate $B$ & strong $B$\\
		\hline
		\hline
		cyl. TI nanowire \cite{zhang2009, egger2010, imura2011, dufouleur2013, tian2013,peng2010,bardarson2010, zhang2010, rosenberg2010, bardarson2013, cho2015, ziegler2018,Kozlovsky2020,zhang2011,vafek2011,sitte2012,brey2014,koenig2014} & osc. (SBO)  & quan. cond. pl. (CSS) & osc. (SBO) & osc. (SBO) & osc. (SBO)\\
		\hline
		TI nanocone \cite{Kozlovsky2020} & cond. steps (MO) & quan. cond. pl. (CSS) & cond. steps (MO) & res. trans. (LL)& CB/curr. supp. (LL) \\
		\hline
		TI dumbbell & cond. steps (MO) & quan. cond. pl. (CSS) & \
		cond. steps (MO) & res. trans./CB/curr. supp. (LL)  & triple QD/curr. supp. (LL)\\
		\hline
	\end{tabular}
	\caption{Summary of low-energy magnetotransport regimes in simple examples of smoothly shaped TI nanowires, with comparison to the cylindrical limit. Each entry describes the shape of $G(E_F)$ for fixed $B$ with its physical origin in brackets. If the transport regime for a given geometry and $B$-field depends on the Fermi level $E_F$, possible different regimes are separated by a slash.  Abbreviations: \textit{osc.}~$\textendash$ Van Hove singularity-induced conductance oscillation on top of an increasing conductance background; \textit{SBO} $\textendash$ subband opening as $E_F$ is increased; \textit{quan.~cond.~pl.} $\textendash$ quantized conductance plateau; \textit{CSS} $\textendash$ chiral side surface states; \textit{cond.~steps} $\textendash$ conductance steps; \textit{MO} $\textendash$ mode opening; \textit{res.~trans.} $\textendash$ resonant transmission; \textit{LL} $\textendash$ Landau levels;  \textit{CB} $\textendash$ Coulomb blockade; \textit{curr.~supp.} $\textendash$ current suppression; \textit{triple QD} $\textendash$ transport through a triple quantum dot.}
	\label{tab1}
\end{table*}


\section{Summary \& Conclusions}
\label{conclusion}

We showed that numerous transport regimes are accessible when a shaped TI nanowire (TINW), 
\ie~an axially symmetric TINW with varying cross section along its length, is immersed in a homogeneous magnetic field.
The results are summarized in Tab.~\ref{tab1}, which is briefly outlined in the following.

Consider first a strong ($l_B \ll$ wire width) perpendicular magnetic field ($\mathbf{B}\perp\hat{z}$), such that the top and bottom TINW surfaces are in the quantum Hall regime. Here, transport is dominated by chiral side surface states which do not ``feel'' the geometry of the nanowire. Thus, shaped TINWs behave qualitatively the same as cylindrical TINWs.
A characteristic transport feature in this regime is a quantized conductance plateau in a magnetic field-dependent energy window \cite{Kozlovsky2020,zhang2011,vafek2011,sitte2012,brey2014,koenig2014}. In contrast, for weak perpendicular magnetic fields, states wrap around the wire and transport is thus geometry-dependent. For cylindrical nanowires, subbands open as $E_F$ is increased, which leads to Van Hove singularity-induced oscillations on top of an increasing conductance background. For shaped TINWs in a weak perpendicular magnetic field, quantum confinement-induced potentials need to be overcome. The corresponding transport signatures are steps in the conductance due to mode opening.
 
In a coaxial magnetic field ($\mathbf{B}\parallel\hat{z}$), the focus of our work, states wrap around the circumference and enclose the magnetic flux -- hence, transport is highly sensitive to both magnetic field and geometry. 
In the presence of rotational symmetry, the angular momentum $l$ is a good quantum number and the effective mass-type potential $|V_l|$ is a useful tool to predict the transport behavior of any shaped TINW. For cylindrical nanowires, $G(E_F)$ is determined by Van Hove singularity-induced oscillations on top of an increasing conductance background, independently of the magnetic field strength -- a result of translational invariance along the wire. Smoothed \emph{TI nanocones (TINCs)} in contrast, first introduced in Ref. \cite{Kozlovsky2020} and discussed at length in Sec.~\ref{TINC}, can be tuned between three different regimes by a simple variation of the coaxial magnetic field strength: conductance steps due to mode opening (a mode with angular momentum quantum number $l$ opens as soon as $E_F > |V_l|$), resonant transmission, and Coulomb blockade-like transport through LLs.

Using the single TINC as a building block, more complex-shaped TINWs can be composed.  We focused on the paradigmatic example of a TINW constriction, dubbed \emph{TI dumbbell}. In this system, a triple quantum dot structure can form for magnetic fields beyond approximately 3-4 Teslas (for our choice of parameters), adding one more fundamentally different transport regime to those available in a TINC.
Most notably however, in the intermediate magnetic field regime ($B\approx1-2\,$T), one can switch on and off Coulomb blockade oscillations at will, simply upon altering the $B$-value:
If $E_F$ is far from the disorder-broadened LL energies, the TINCs on the sides of the dumbbell act as infinitely strong barriers and transport is suppressed.  On the other hand, the same TINCs act as finite tunnel barriers if $E_F$ is in the vicinity of (but not exactly in resonance with) the LL energies, so that conductance oscillations of Coulomb blockade type should be visible if a gate electrode is applied to the central region.

Concerning experimental realization, we point out that TI nanowires with non-uniform cross section have been fabricated, 
see \eg Refs. \cite{Kessel_2017, ziegler2018}.  All-round (homogeneous) gating, as assumed in the discussion of Coulomb blockade, is also currently possible \cite{Storm2012,Royo2017}.  The realization of a finely shaped TI tube thus appears challenging but within current experimental capabilities. Moreover, shaped TINWs represent a substantial practical advantage from the experimental point of view:
The Coulomb blockade regime in the TI dumbbell arises from magnetic confinement of Dirac electrons.  To the contrary of proposals for 2D geometries, where non-homogeneous magnetic fields are necessary to confine Dirac fermions \cite{demartino2007,masir2008}, in our shaped nanowires a homogeneous magnetic field is enough (see also Appendix \ref{AppB}). This fact is the decisive ingredient behind the relatively simple on/off-switch mechanism between different magnetotransport regimes.

Finally, let us stress that magnetic confinement is not restricted to the axially symmetric TI dumbbell explicitly considered.  Essentially, one can use TINCs -- tunable into barriers due to Landau quantization -- and cylindrical TINWs -- where free motion follows from $B_\perp=0$ -- as elemental building blocks and connect them in series to build arbitrary magnetic barrier profiles in homogeneous magnetic fields.  Furthermore, strict axial symmetry is not required: a TINC with a somewhat distorted cross-section acts as a magnetic barrier as long as the local $l_B$ is smaller than its geometrical size, ensuring the formation of QH states throughout its (distorted) perimeter.
We thus expect our results to be valid guidelines for the analysis of magnetotransport in a wide range of TINWs of any shape. 

{\em Acknowledgements:}  
We thank Andrea Donarini, Milena Grifoni, Dominik Hahn, Gilles Montambaux and Max Nitsch for useful discussions. CG thanks the STherQO members, and in particular Guillaume Weick, for useful comments. This work was funded by the Deutsche Forschungsgemeinschaft (DFG, German Research
Foundation) Project-ID 314695032—SFB 1277 (subproject A07), 
and within Priority Programme SPP 1666 "Topological Insulators" (project Ri681-12/2). Support
by the Elitenetzwerk Bayern Doktorandenkolleg "Topological Insulators" as well as the \'Ecole Doctorale Physique en \^Ile de France (EDPIF) is also acknowledged.

\appendix
\section{Dirac surface Hamiltonian for a shaped TI nanowire}
\label{AppA}

The derivation of the Hamiltonian \eqref{mainham} using a fiel theoretic approach is sketched here very briefly; the details can be found in Ref.~[\onlinecite{xypakis2017}].

Fermions on a (2+1)-dimensional curved manifold fulfill the covariant Dirac equation
\begin{equation}
\gamma^\mu D_\mu\Psi=0,
\label{dirac}
\end{equation} 
where $D_\mu\equiv\partial_\mu+\Gamma_\mu$ is the covariant derivative and $\gamma^\mu\equiv V_a^\mu\xi^a$ are covariant Dirac matrices. The additional term $\Gamma_\mu$ is known as the spin connection \cite{fecko2006,koke2016}. The $\xi^a$ are local Dirac matrices satisfying the Clifford algebra $\{\xi^a,\xi^b\}=2\eta^{ab}$, where $\eta^{ab}$ is the Minkowski metric, and $V_a^\mu$ are the inverse vielbeins.

The metric for the shaped TINW is given as \cite{xypakis2017} $dl^2=-dt^2+(1+R'^2)dz^2+R^2d\varphi^2$, where $\varphi$ is the azimuthal angle and $R\equiv R(z)$ is the radius as a function of the coaxial coordinate $z$; note that we work in natural units. 
We choose the following set of Dirac matrices (different from the choice in Ref. \cite{xypakis2017}):
\begin{equation}
\xi^0=i\sigma_x,\hspace{0.5cm}\xi^1=\sigma_y,\hspace{0.5cm}\xi^2=-\sigma_z,
\end{equation}
such that
\begin{equation}
\gamma^0=i\sigma_x,\hspace{0.5cm}\gamma^1=\frac{1}{\sqrt{1+R'^2}}\sigma_y,\hspace{0.5cm}\gamma^2=-\frac{1}{R}\sigma_z.
\end{equation}
For the spin connection one finds $\Gamma_t=\Gamma_z=0$ and 
\begin{equation}
\Gamma_\varphi=\frac{i}{2}\frac{R'}{\sqrt{1+R'^2}}\sigma_x.
\end{equation}
Then the Dirac equation (\ref{dirac}) becomes
\begin{equation}
\begin{aligned}
i\sigma_x\partial_t\Psi=&\left[-\frac{1}{\sqrt{1+R'^2}}\left(\partial_z+\frac{R'}{2R}\right)\sigma_y+\frac{1}{R}\partial_\varphi\sigma_z\right]\Psi.
\end{aligned}
\end{equation}
Restoring the fundamental constants and left-multiplying by $\sigma_x$ such that a Hamiltonian can be defined by $H\Psi=i\hbar\partial_t\Psi$, one arrives at the surface Dirac Hamiltonian for a shaped TINW:
\begin{equation}
H=v_F\left[\frac{1}{\sqrt{1+R'^2}}\left(p_z-\frac{i\hbar}{2}\frac{R'}{R}\right)\sigma_z+p_\varphi\sigma_y\right],
\label{xypakis}
\end{equation}
where $p_z\equiv-i\hbar\partial_z$ and $p_\varphi\equiv-i\hbar R^{-1}\partial_\varphi$. The term $\propto R'/R$ represents the nontrivial spin connection.

In the presence of a homogeneous coaxial magnetic field $\mathbf{B}=B\hat{z}$, the vector potential in the symmetric gauge is given by $\mathbf{A}=A_\varphi\,\hat{\varphi}=Br/2\,\hat{\varphi}$.
We replace $p_\varphi\rightarrow p_\varphi+eA_\varphi$, where $e>0$, to obtain
\begin{equation}
\begin{aligned}
H=v_F&\left[\frac{1}{\sqrt{1+R'^2}}\left(p_z-\frac{i\hbar}{2}\frac{R'}{R}\right)\sigma_z\right.\\&\left.+\left(p_\varphi+\frac{\hbar}{R}\frac{\Phi}{\Phi_0}\right)\sigma_y\right].
\label{xypakis2}
\end{aligned}
\end{equation}
Here, $\Phi\equiv \pi BR^2$ is the magnetic flux piercing the wire and $\Phi_0\equiv h/e$ is the magnetic flux quantum. This is the Hamiltonian (\ref{mainham}) provided in the main part of this paper.

\section{Conductance simulations with \textsc{KWANT}}
\label{AppConductanceKwant}
	\textit{Non-uniform lattice --}
	The effective tight-binding Hamiltonian used to compute transport throughout this work with the software package \textsc{kwant} is obtained by discretizing Eq.~\eqref{ourHam}.
	In the following, we use the short-hand notation $\Psi(s_i, \varphi_j) \equiv \Psi_{i,j}$ for the two-component spinor wave function $\Psi$ on the numerical grid defined by the grid points $(i,j)$ (where $i,j$ are integers and $s$ is the arclength along the wire).
	Using this notation, a discretization of the transversal wave number operator $\hat{k}_\varphi=-\ci \partial_\varphi /R(s)$ with the standard symmetric finite difference method yields
	\begin{align}
	\hat{k}_\varphi(s) \Psi_{i,j}  \rightarrow &- \frac{\ci}{R(s_i)} \frac{1}{2 \Delta \varphi} \left(\Psi_{i,j+1} - \Psi_{i,j-1} \right) \\
	\equiv &- \frac{\ci}{2a_\varphi(s_i)} \left(\Psi_{i,j+1} - \Psi_{i,j-1} \right),
	\label{eq:cone_momentum_discretized}
	\end{align}
	where the angle $\Delta \varphi$ is determined by the number of grid points in the transversal direction $N_\varphi$, namely $\Delta \varphi= 2\pi/N_\varphi$. In Eq.~\eqref{eq:cone_momentum_discretized}, we introduce the $s$-dependent transversal grid constant $a_\varphi(s) \equiv R(s) \Delta \varphi$ to highlight that effectively the transversal grid spacing is changing such that $N_\varphi a_\varphi(s) = 2 \pi R(s)$.

	With the standard discretization of the longitudinal wave number operator $\hat{k}_s  \Psi_{i,j} = -\ci \left(\Psi_{i+1,j} - \Psi_{i-1,j} \right)/(2 a_s)$, where $a_s$ is the grid spacing in the longitudinal direction, we arrive at the tight-binding\index{tight-binding} Hamiltonian 
	\begin{align}
	\begin{split}
	H_\text{TB} = - \frac{\ci \hbar v_F}{2} \sum_{i,j} \left( \frac{1}{a_s} \sigma_z \ket{i,j}\bra{i+1,j}
	\right. \\
	+ \left. \frac{1}{a_\varphi(s_i)} \sigma_y \ket{i,j}\bra{i,j+1} \right)  + \text{h.c.}
	\end{split}
	\label{eq:tight_binding_hamiltonian}
	\end{align}
	The coaxial magnetic field is implemented using the usual Peierls substitution. 

	\textit{Modeling disorder on curved surfaces --}
	For creating correlated disorder we use the so-called Fourier filtering method (FFM), which is discussed in detail for instance in Ref. \cite{Zierenberg2017}. 
	For shaped TINWs, we construct a disorder landscape with the desired correlation length in a 3D box, in which the TINW is embedded. The values for the disorder potential $V_\text{dis}(\mathbf{r})$ are then evaluated within the box on the surface of the TINW and added as an onsite potential to the tight-binding Hamiltonian \eqref{eq:tight_binding_hamiltonian}.

\section{Effective mass potential for graphene subject to a magnetic step barrier}
\label{AppB}

In view of the step-like profile of $B_\perp$ for a single TINC, see Fig.~\ref{fig_cones}(a), we here provide the connection to the related and well-known problem of a magnetic step barrier in (single-valley) graphene \cite{demartino2007}. More generally, the form of Eq.~(\ref{laplaceeqq}) that we found for a shaped TINW is very similar to the effective Schr\"odinger equation found in graphene subject to various magnetic field profiles \cite{demartino2007,masir2008,ghosh2008,roy2012}. In this Appendix, we show that the results of Ref. \cite{demartino2007} can be reinterpreted in the language of an effective mass potential, in full analogy to the effective potential introduced in Eq.~(\ref{effpot}). 

Consider an infinite graphene sheet, subject to a magnetic step barrier that is translationally invariant in the $y$-direction and nonzero only in the region $-d\leq x\leq d$, such that $B(x,y)=B_0\Theta(d^2-x^2)$ \cite{demartino2007}. Assume inter-valley scattering to be absent. Choosing the gauge $\mathbf{A}(x,y)=A(x)\hat{y}$, where
\begin{equation}
A(x)=B_0\begin{cases}-d, & x<-d \,\, (\text{region I})\\
x, & |x|\leq d \,\, (\text{region II})\\
d, & x>d \,\, (\text{region III})
\end{cases}
\label{gaugeg}
\end{equation}
the Dirac equation becomes
\begin{equation}
v_F\left\{p_x\sigma_x+\left[\hbar k_y+eA(x)\right]\sigma_y\right\}\psi(x)=\epsilon\psi(x),
\label{woasned}
\end{equation}
where we exploited the fact that transverse momentum $\hbar k_y$ is a good quantum number. Equation (\ref{woasned}) is easily decoupled to give
\begin{equation}
\mathcal{O}_y^\pm\psi_\pm=\epsilon^2\psi_\pm,
\label{grapheq}
\end{equation}
where the Dirac spinor $\psi=(\psi_+,\psi_-)^T$ and
\begin{equation}
\begin{aligned}
\mathcal{O}_y^\pm&=-(\hbar v_F)^2(\partial_x^2+\mathcal{P}_y^\pm),\\
\mathcal{P}_y^\pm&=\mp\frac{e}{\hbar}A'-\left(k_y+\frac{e}{\hbar}A\right)^2.
\end{aligned}
\end{equation}
The analogy to Eq. (\ref{laplaceeqq}) is evident, discrete angular momentum being replaced by continuous transverse momentum, and the coaxial coordinate replaced by $x$. Comparing Eq.~(\ref{woasned}) to Eq.~(\ref{Direq}), we observe that transverse momentum acts as a mass potential: $V(x)\equiv\hbar v_Fk(x)=\hbar v_F(k_y+eA(x)/\hbar)$. With the gauge (\ref{gaugeg}) one has \cite{demartino2007}
\begin{equation}
k(x)=\tilde{\epsilon}\sin\phi+\begin{cases}0, & x<-d \,\, (\text{region I})\\
(d+x)/l_B^2, & |x|\leq d \,\, (\text{region II})\\
2d/l_B^2, & x>d \,\, (\text{region III})
\end{cases}
\label{kgraph}
\end{equation}
where $\tilde{\epsilon}\equiv\epsilon/(\hbar v_F)$, $l_B\equiv\sqrt{\hbar/(eB_0)}$ is the magnetic length and $\phi$ is the kinematic incidence angle.
Now, if $\sin\phi\geq 0$, it is clear that $k(x)\geq0$, \ie, $V(x)$ cannot become negative. However, if $\sin\phi<0$, we have two possibilities: (i) $\tilde{\epsilon}|\sin\phi|\geq2d/l_B^2$, then $k(x)\leq0$ always. (ii) $\tilde{\epsilon}|\sin\phi|<2d/l_B^2$, then
\begin{equation}
k(x)\begin{cases}<0, & x<-d \,\, (\text{region I})\\
\leq0, & -d\leq x\leq x_0 \,\, (\text{region II})\\
>0, & x_0<x\leq d \,\, (\text{region II})\\
>0. & x>d \,\, (\text{region III})\\
\end{cases}
\end{equation}
Here, $x_0\equiv\tilde{\epsilon}|\sin\phi|l_B^2-d<d$ denotes the root of the effective potential, analogous to $\tilde{z}_l$, cf.~Eq.~(\ref{root}).
In full analogy to Section \ref{TINCpot}, a root in $V(x)$ corresponds to a minimum and a surrounding potential wedge in $|V(x)|$.

\begin{figure}
	\centering
	\includegraphics[width=.9\columnwidth]{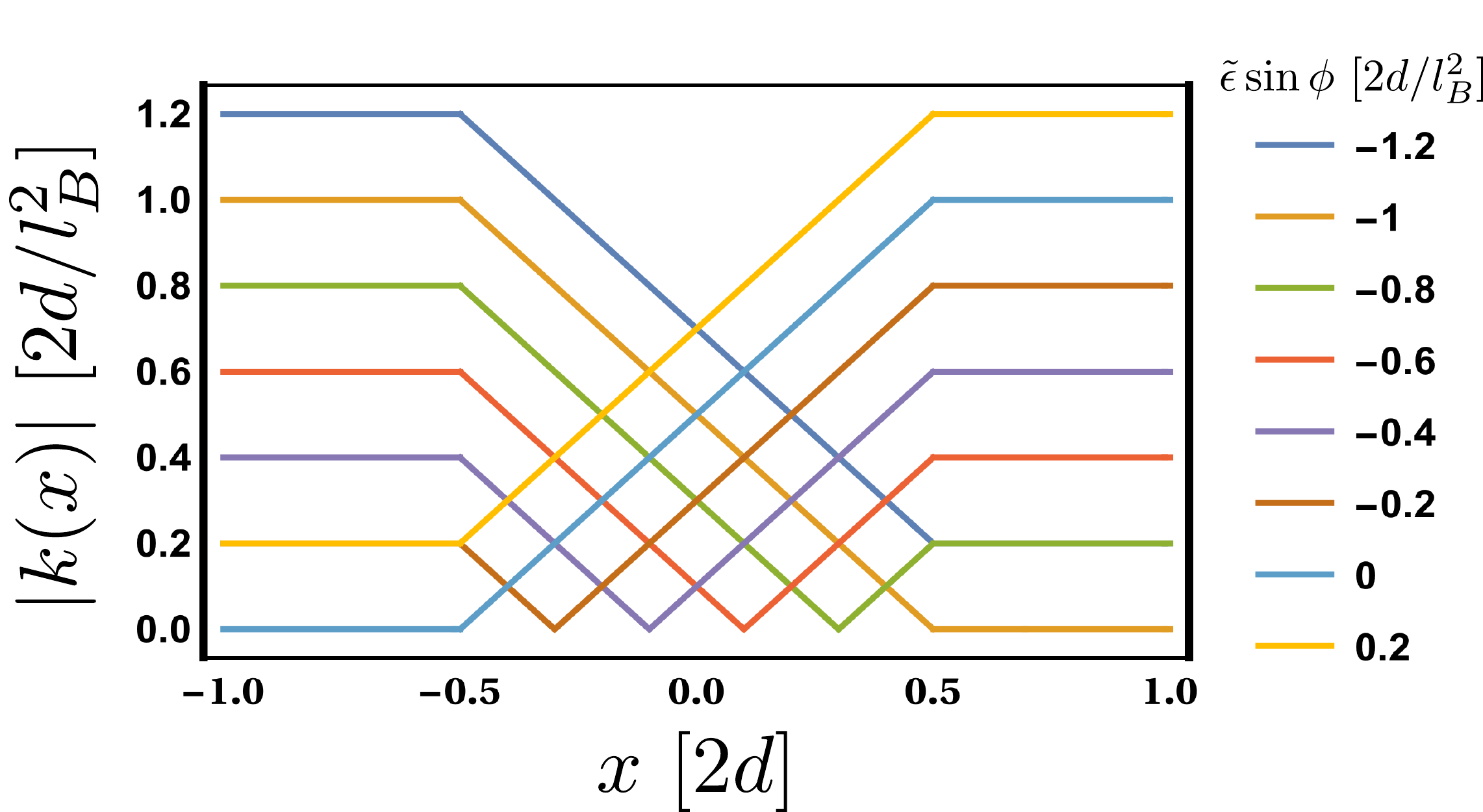}
	\caption{Effective potential landscape $|k(x)|$ seen by a state incident on a magnetic step barrier in graphene, for different values of transverse momentum ($d$ and $l_B$ are fixed), and as calculated from Eq.~(\ref{kgraph}). The linear form of the vector potential leads to perfectly triangular potential wells.}
	\label{fig:grapheneabsVeff}
\end{figure}

Consequently, bound states within the effective potential $|V(x)|$ may exist if the two necessary criteria $\sin\phi<0$ and $|\sin\phi|<2d/(\tilde{\epsilon}l_B^2)$ are fulfilled.
This is visualized in Fig.~\ref{fig:grapheneabsVeff}.
If bound states exist, they correspond to Landau levels (LLs) \cite{masir2008}. This duality of Landau level formation and bound states in the effective potential at LL energies is in complete analogy to what we find in Section \ref{TINCstruc}.

Moreover, the picture of an effective potential landscape $|V(x)|$ can explain intuitively the perfect reflection criterion found in Ref. \cite{demartino2007}, Eq. (12):
\begin{equation}
\tilde{\epsilon}\leq d/l_B^2.
\label{crit}
\end{equation}
When considering Fig.~\ref{fig:grapheneabsVeff} and varying the parameter $\tilde{\epsilon}\sin\phi$ arbitrarily, it is clear that the minimal energy threshold $k_\text{max}\equiv\text{max}(|k(-d)|,|k(d)|)$ an incoming state can see (the analog of $\epsilon_l^\text{max}$ in Section \ref{TINCstruc}) is $k_\text{max}=d/l_B^2$. Hence, no transmission can occur in principle if $\tilde{\epsilon}\leq k_\text{max}$. This is precisely the criterion (\ref{crit}).

Obviously, the discussion conducted in this Appendix for a simple magnetic step barrier can be extended to situations where more complicated inhomogeneous magnetic fields profiles are applied to graphene. For example, for each of the magnetic field profiles studied in Ref. \cite{masir2008}, we can construct the corresponding shaped TINW by choice of the profile of $B_\perp(z)$, cf. the insets of Fig.~\ref{fig_cones}. 

\section{Parametrization of shaped TI nanowires}
\label{Appsmoothcone}
We use the \textit{NDEigensystem} routine in \textit{Wolfram Mathematica} to solve Eq.~\eqref{laplaceeqq} numerically.
The TI nanocone is parametrized with
\begin{equation}
\begin{aligned}
R_\sigma^\text{TINC}(z)&\equiv R_0+(R_1-R_0)\Theta_\sigma(z-z_1)\\
&+\mathcal{S}(z-z_0)[\Theta_\sigma(z-z_0)-\Theta_\sigma(z-z_1)],
\label{coneapprox}
\end{aligned}
\end{equation}
where $
\Theta_\sigma(z-z')\equiv \frac{1}{2}+\frac{1}{\pi}\arctan[\sigma(z-z')]$
is a smoothed Heaviside function with a step at $z=z'$, such that $\Theta_{\sigma\to\infty}(z-z')=\Theta(z-z')$. 
With this, one can construct any shaped TINW at will. For example, for the TI dumbbell, see Fig.~\ref{fig_cones}, where we assume $z_3-z_2=z_1-z_0$ and $R_2=R_1$, $R_3=R_0$ for simplicity, we have 
\begin{equation}
R_\sigma^\text{TIDB}(z)\equiv\, R_\sigma^\text{TINC}(z)+R_\sigma^\text{TINC}(-z+z_1+z_2)-R_1. 
\end{equation}
Changing the value of $\sigma$ allows to interpolate between the ideal TI dumbbell [Fig. \ref{fig_cones}(b)] and the smoothed version shown in Fig. \ref{fig_cones}(c).

\bibliography{ref2}

\begin{thebibliography}{52}%
\makeatletter
\providecommand \@ifxundefined [1]{%
 \@ifx{#1\undefined}
}%
\providecommand \@ifnum [1]{%
 \ifnum #1\expandafter \@firstoftwo
 \else \expandafter \@secondoftwo
 \fi
}%
\providecommand \@ifx [1]{%
 \ifx #1\expandafter \@firstoftwo
 \else \expandafter \@secondoftwo
 \fi
}%
\providecommand \natexlab [1]{#1}%
\providecommand \enquote  [1]{``#1''}%
\providecommand \bibnamefont  [1]{#1}%
\providecommand \bibfnamefont [1]{#1}%
\providecommand \citenamefont [1]{#1}%
\providecommand \href@noop [0]{\@secondoftwo}%
\providecommand \href [0]{\begingroup \@sanitize@url \@href}%
\providecommand \@href[1]{\@@startlink{#1}\@@href}%
\providecommand \@@href[1]{\endgroup#1\@@endlink}%
\providecommand \@sanitize@url [0]{\catcode `\\12\catcode `\$12\catcode
  `\&12\catcode `\#12\catcode `\^12\catcode `\_12\catcode `\%12\relax}%
\providecommand \@@startlink[1]{}%
\providecommand \@@endlink[0]{}%
\providecommand \url  [0]{\begingroup\@sanitize@url \@url }%
\providecommand \@url [1]{\endgroup\@href {#1}{\urlprefix }}%
\providecommand \urlprefix  [0]{URL }%
\providecommand \Eprint [0]{\href }%
\providecommand \doibase [0]{http://dx.doi.org/}%
\providecommand \selectlanguage [0]{\@gobble}%
\providecommand \bibinfo  [0]{\@secondoftwo}%
\providecommand \bibfield  [0]{\@secondoftwo}%
\providecommand \translation [1]{[#1]}%
\providecommand \BibitemOpen [0]{}%
\providecommand \bibitemStop [0]{}%
\providecommand \bibitemNoStop [0]{.\EOS\space}%
\providecommand \EOS [0]{\spacefactor3000\relax}%
\providecommand \BibitemShut  [1]{\csname bibitem#1\endcsname}%
\let\auto@bib@innerbib\@empty
\bibitem [{\citenamefont {Kozlovsky}\ \emph {et~al.}(2020)\citenamefont
  {Kozlovsky}, \citenamefont {Graf}, \citenamefont {Kochan}, \citenamefont
  {Richter},\ and\ \citenamefont {Gorini}}]{Kozlovsky2020}%
  \BibitemOpen
  \bibfield  {author} {\bibinfo {author} {\bibfnamefont {Raphael}\ \bibnamefont
  {Kozlovsky}}, \bibinfo {author} {\bibfnamefont {Ansgar}\ \bibnamefont
  {Graf}}, \bibinfo {author} {\bibfnamefont {Denis}\ \bibnamefont {Kochan}},
  \bibinfo {author} {\bibfnamefont {Klaus}\ \bibnamefont {Richter}}, \ and\
  \bibinfo {author} {\bibfnamefont {Cosimo}\ \bibnamefont {Gorini}},\
  }\bibfield  {title} {\enquote {\bibinfo {title} {{Magnetoconductance, Quantum
  Hall Effect, and Coulomb Blockade in Topological Insulator Nanocones}},}\
  }\href {\doibase 10.1103/PhysRevLett.124.126804} {\bibfield  {journal}
  {\bibinfo  {journal} {Phys. Rev. Lett.}\ }\textbf {\bibinfo {volume} {124}},\
  \bibinfo {pages} {126804} (\bibinfo {year} {2020})}\BibitemShut {NoStop}%
\bibitem [{\citenamefont {Bansil}\ \emph {et~al.}(2016)\citenamefont {Bansil},
  \citenamefont {Lin},\ and\ \citenamefont {Das}}]{bansil2016}%
  \BibitemOpen
  \bibfield  {author} {\bibinfo {author} {\bibfnamefont {A.}~\bibnamefont
  {Bansil}}, \bibinfo {author} {\bibfnamefont {Hsin}\ \bibnamefont {Lin}}, \
  and\ \bibinfo {author} {\bibfnamefont {Tanmoy}\ \bibnamefont {Das}},\
  }\bibfield  {title} {\enquote {\bibinfo {title} {\emph{Colloquium}:
  Topological band theory},}\ }\href {\doibase 10.1103/RevModPhys.88.021004}
  {\bibfield  {journal} {\bibinfo  {journal} {Rev. Mod. Phys.}\ }\textbf
  {\bibinfo {volume} {88}},\ \bibinfo {pages} {021004} (\bibinfo {year}
  {2016})}\BibitemShut {NoStop}%
\bibitem [{\citenamefont {Hasan}\ and\ \citenamefont {Kane}(2010)}]{hasan2010}%
  \BibitemOpen
  \bibfield  {author} {\bibinfo {author} {\bibfnamefont {M.~Z.}\ \bibnamefont
  {Hasan}}\ and\ \bibinfo {author} {\bibfnamefont {C.~L.}\ \bibnamefont
  {Kane}},\ }\bibfield  {title} {\enquote {\bibinfo {title} {\emph{Colloquium}:
  Topological insulators},}\ }\href {\doibase 10.1103/RevModPhys.82.3045}
  {\bibfield  {journal} {\bibinfo  {journal} {Rev. Mod. Phys.}\ }\textbf
  {\bibinfo {volume} {82}},\ \bibinfo {pages} {3045--3067} (\bibinfo {year}
  {2010})}\BibitemShut {NoStop}%
\bibitem [{\citenamefont {Ando}(2013)}]{ando2013}%
  \BibitemOpen
  \bibfield  {author} {\bibinfo {author} {\bibfnamefont {Yoichi}\ \bibnamefont
  {Ando}},\ }\bibfield  {title} {\enquote {\bibinfo {title} {{Topological
  Insulator Materials}},}\ }\href {\doibase 10.7566/JPSJ.82.102001} {\bibfield
  {journal} {\bibinfo  {journal} {Journal of the Physical Society of Japan}\
  }\textbf {\bibinfo {volume} {82}},\ \bibinfo {pages} {102001} (\bibinfo
  {year} {2013})}\BibitemShut {NoStop}%
\bibitem [{\citenamefont {Zhang}\ \emph
  {et~al.}(2009{\natexlab{a}})\citenamefont {Zhang}, \citenamefont {Liu},
  \citenamefont {Qi}, \citenamefont {Dai}, \citenamefont {Fang},\ and\
  \citenamefont {Zhang}}]{SCzhang2009}%
  \BibitemOpen
  \bibfield  {author} {\bibinfo {author} {\bibfnamefont {Haijun}\ \bibnamefont
  {Zhang}}, \bibinfo {author} {\bibfnamefont {Chao-Xing}\ \bibnamefont {Liu}},
  \bibinfo {author} {\bibfnamefont {Xiao-Liang}\ \bibnamefont {Qi}}, \bibinfo
  {author} {\bibfnamefont {Xi}~\bibnamefont {Dai}}, \bibinfo {author}
  {\bibfnamefont {Zhong}\ \bibnamefont {Fang}}, \ and\ \bibinfo {author}
  {\bibfnamefont {Shou-Cheng}\ \bibnamefont {Zhang}},\ }\bibfield  {title}
  {\enquote {\bibinfo {title} {{Topological insulators in
  ${\mathrm{Bi}}_{2}{\mathrm{Se}}_{3}$, ${\mathrm{Bi}}_{2}{\mathrm{Te}}_{3}$
  and ${\mathrm{Sb}}_{2}{\mathrm{Te}}_{3}$ with a single Dirac cone on the
  surface}},}\ }\href {https://doi.org/10.1038/nphys1270} {\bibfield  {journal}
  {\bibinfo  {journal} {Nature Physics}\ }\textbf {\bibinfo {volume} {5}},\
  \bibinfo {pages} {438--442} (\bibinfo {year}
  {2009}{\natexlab{a}})}\BibitemShut {NoStop}%
\bibitem [{Note1()}]{Note1}%
  \BibitemOpen
  \bibinfo {note} {See {\protect \it e.g.~}Refs.~\cite
  {tokoyama2010,schwab2011,lee2015} or Refs.~\cite {culcer2012,ando2017} for
  reviews.}\BibitemShut {Stop}%
\bibitem [{\citenamefont {Zhang}\ \emph
  {et~al.}(2009{\natexlab{b}})\citenamefont {Zhang}, \citenamefont {Ran},\ and\
  \citenamefont {Vishwanath}}]{zhang2009}%
  \BibitemOpen
  \bibfield  {author} {\bibinfo {author} {\bibfnamefont {Yi}~\bibnamefont
  {Zhang}}, \bibinfo {author} {\bibfnamefont {Ying}\ \bibnamefont {Ran}}, \
  and\ \bibinfo {author} {\bibfnamefont {Ashvin}\ \bibnamefont {Vishwanath}},\
  }\bibfield  {title} {\enquote {\bibinfo {title} {{Topological insulators in
  three dimensions from spontaneous symmetry breaking}},}\ }\href {\doibase
  10.1103/PhysRevB.79.245331} {\bibfield  {journal} {\bibinfo  {journal} {Phys.
  Rev. B}\ }\textbf {\bibinfo {volume} {79}},\ \bibinfo {pages} {245331}
  (\bibinfo {year} {2009}{\natexlab{b}})}\BibitemShut {NoStop}%
\bibitem [{\citenamefont {Egger}\ \emph {et~al.}(2010)\citenamefont {Egger},
  \citenamefont {Zazunov},\ and\ \citenamefont {Yeyati}}]{egger2010}%
  \BibitemOpen
  \bibfield  {author} {\bibinfo {author} {\bibfnamefont {R.}~\bibnamefont
  {Egger}}, \bibinfo {author} {\bibfnamefont {A.}~\bibnamefont {Zazunov}}, \
  and\ \bibinfo {author} {\bibfnamefont {A.~Levy}\ \bibnamefont {Yeyati}},\
  }\bibfield  {title} {\enquote {\bibinfo {title} {{Helical Luttinger Liquid in
  Topological Insulator Nanowires}},}\ }\href {\doibase
  10.1103/PhysRevLett.105.136403} {\bibfield  {journal} {\bibinfo  {journal}
  {Phys. Rev. Lett.}\ }\textbf {\bibinfo {volume} {105}},\ \bibinfo {pages}
  {136403} (\bibinfo {year} {2010})}\BibitemShut {NoStop}%
\bibitem [{\citenamefont {Imura}\ \emph {et~al.}(2011)\citenamefont {Imura},
  \citenamefont {Takane},\ and\ \citenamefont {Tanaka}}]{imura2011}%
  \BibitemOpen
  \bibfield  {author} {\bibinfo {author} {\bibfnamefont {Ken-Ichiro}\
  \bibnamefont {Imura}}, \bibinfo {author} {\bibfnamefont {Yositake}\
  \bibnamefont {Takane}}, \ and\ \bibinfo {author} {\bibfnamefont {Akihiro}\
  \bibnamefont {Tanaka}},\ }\bibfield  {title} {\enquote {\bibinfo {title}
  {{Spin Berry phase in anisotropic topological insulators}},}\ }\href
  {\doibase 10.1103/PhysRevB.84.195406} {\bibfield  {journal} {\bibinfo
  {journal} {Phys. Rev. B}\ }\textbf {\bibinfo {volume} {84}},\ \bibinfo
  {pages} {195406} (\bibinfo {year} {2011})}\BibitemShut {NoStop}%
\bibitem [{\citenamefont {Dufouleur}\ \emph {et~al.}(2013)\citenamefont
  {Dufouleur}, \citenamefont {Veyrat}, \citenamefont {Teichgr{\"a}ber},
  \citenamefont {Neuhaus}, \citenamefont {Nowka}, \citenamefont {Hampel},
  \citenamefont {Cayssol}, \citenamefont {Schumann}, \citenamefont {Eichler},
  \citenamefont {Schmidt}, \citenamefont {B{\"u}chner},\ and\ \citenamefont
  {Giraud}}]{dufouleur2013}%
  \BibitemOpen
  \bibfield  {author} {\bibinfo {author} {\bibfnamefont {J.}~\bibnamefont
  {Dufouleur}}, \bibinfo {author} {\bibfnamefont {L.}~\bibnamefont {Veyrat}},
  \bibinfo {author} {\bibfnamefont {A.}~\bibnamefont {Teichgr{\"a}ber}},
  \bibinfo {author} {\bibfnamefont {S.}~\bibnamefont {Neuhaus}}, \bibinfo
  {author} {\bibfnamefont {C.}~\bibnamefont {Nowka}}, \bibinfo {author}
  {\bibfnamefont {S.}~\bibnamefont {Hampel}}, \bibinfo {author} {\bibfnamefont
  {J.}~\bibnamefont {Cayssol}}, \bibinfo {author} {\bibfnamefont
  {J.}~\bibnamefont {Schumann}}, \bibinfo {author} {\bibfnamefont
  {B.}~\bibnamefont {Eichler}}, \bibinfo {author} {\bibfnamefont {O.~G.}\
  \bibnamefont {Schmidt}}, \bibinfo {author} {\bibfnamefont {B.}~\bibnamefont
  {B{\"u}chner}}, \ and\ \bibinfo {author} {\bibfnamefont {R.}~\bibnamefont
  {Giraud}},\ }\bibfield  {title} {\enquote {\bibinfo {title} {{Quasiballistic
  Transport of Dirac Fermions in a ${\mathrm{Bi}}_{2}{\mathrm{Se}}_{3}$
  Nanowire}},}\ }\href {\doibase 10.1103/PhysRevLett.110.186806} {\bibfield
  {journal} {\bibinfo  {journal} {Phys. Rev. Lett.}\ }\textbf {\bibinfo
  {volume} {110}},\ \bibinfo {pages} {186806} (\bibinfo {year}
  {2013})}\BibitemShut {NoStop}%
\bibitem [{\citenamefont {Tian}\ \emph {et~al.}(2013)\citenamefont {Tian},
  \citenamefont {Ning}, \citenamefont {Qu}, \citenamefont {Du}, \citenamefont
  {Wang},\ and\ \citenamefont {Zhang}}]{tian2013}%
  \BibitemOpen
  \bibfield  {author} {\bibinfo {author} {\bibfnamefont {Mingliang}\
  \bibnamefont {Tian}}, \bibinfo {author} {\bibfnamefont {Wei}\ \bibnamefont
  {Ning}}, \bibinfo {author} {\bibfnamefont {Zhe}\ \bibnamefont {Qu}}, \bibinfo
  {author} {\bibfnamefont {Haifeng}\ \bibnamefont {Du}}, \bibinfo {author}
  {\bibfnamefont {Jian}\ \bibnamefont {Wang}}, \ and\ \bibinfo {author}
  {\bibfnamefont {Yuheng}\ \bibnamefont {Zhang}},\ }\bibfield  {title}
  {\enquote {\bibinfo {title} {{Dual evidence of surface Dirac states in thin
  cylindrical topological insulator ${\mathrm{Bi}}_{2}{\mathrm{Te}}_{3}$
  nanowires}},}\ }\href {https://doi.org/10.1038/srep01212} {\bibfield
  {journal} {\bibinfo  {journal} {Scientific Reports}\ }\textbf {\bibinfo
  {volume} {3}},\ \bibinfo {pages} {1212} (\bibinfo {year} {2013})}\BibitemShut
  {NoStop}%
\bibitem [{\citenamefont {Peng}\ \emph {et~al.}(2010)\citenamefont {Peng},
  \citenamefont {Lai}, \citenamefont {Kong}, \citenamefont {Meister},
  \citenamefont {Chen}, \citenamefont {Qi}, \citenamefont {Zhang},
  \citenamefont {Shen},\ and\ \citenamefont {Cui}}]{peng2010}%
  \BibitemOpen
  \bibfield  {author} {\bibinfo {author} {\bibfnamefont {Hailin}\ \bibnamefont
  {Peng}}, \bibinfo {author} {\bibfnamefont {Keji}\ \bibnamefont {Lai}},
  \bibinfo {author} {\bibfnamefont {Desheng}\ \bibnamefont {Kong}}, \bibinfo
  {author} {\bibfnamefont {Stefan}\ \bibnamefont {Meister}}, \bibinfo {author}
  {\bibfnamefont {Yulin}\ \bibnamefont {Chen}}, \bibinfo {author}
  {\bibfnamefont {Xiao-Liang}\ \bibnamefont {Qi}}, \bibinfo {author}
  {\bibfnamefont {Shou-Cheng}\ \bibnamefont {Zhang}}, \bibinfo {author}
  {\bibfnamefont {Zhi-Xun}\ \bibnamefont {Shen}}, \ and\ \bibinfo {author}
  {\bibfnamefont {Yi}~\bibnamefont {Cui}},\ }\bibfield  {title} {\enquote
  {\bibinfo {title} {{Aharonov-Bohm interference in topological insulator
  nanoribbons}},}\ }\href {https://doi.org/10.1038/nmat2609} {\bibfield
  {journal} {\bibinfo  {journal} {Nature Materials}\ }\textbf {\bibinfo
  {volume} {9}},\ \bibinfo {pages} {225--229} (\bibinfo {year}
  {2010})}\BibitemShut {NoStop}%
\bibitem [{\citenamefont {Ostrovsky}\ \emph {et~al.}(2010)\citenamefont
  {Ostrovsky}, \citenamefont {Gornyi},\ and\ \citenamefont
  {Mirlin}}]{ostrovsky2010}%
  \BibitemOpen
  \bibfield  {author} {\bibinfo {author} {\bibfnamefont {P.~M.}\ \bibnamefont
  {Ostrovsky}}, \bibinfo {author} {\bibfnamefont {I.~V.}\ \bibnamefont
  {Gornyi}}, \ and\ \bibinfo {author} {\bibfnamefont {A.~D.}\ \bibnamefont
  {Mirlin}},\ }\bibfield  {title} {\enquote {\bibinfo {title}
  {{Interaction-Induced Criticality in $\mathbb{Z}_2$ Topological
  Insulators}},}\ }\href
  {https://journals.aps.org/prl/abstract/10.1103/PhysRevLett.105.036803}
  {\bibfield  {journal} {\bibinfo  {journal} {Phys. Rev. Lett.}\ }\textbf
  {\bibinfo {volume} {105}},\ \bibinfo {pages} {036803} (\bibinfo {year}
  {2010})}\BibitemShut {NoStop}%
\bibitem [{\citenamefont {Bardarson}\ \emph {et~al.}(2010)\citenamefont
  {Bardarson}, \citenamefont {Brouwer},\ and\ \citenamefont
  {Moore}}]{bardarson2010}%
  \BibitemOpen
  \bibfield  {author} {\bibinfo {author} {\bibfnamefont {J.~H.}\ \bibnamefont
  {Bardarson}}, \bibinfo {author} {\bibfnamefont {P.~W.}\ \bibnamefont
  {Brouwer}}, \ and\ \bibinfo {author} {\bibfnamefont {J.~E.}\ \bibnamefont
  {Moore}},\ }\bibfield  {title} {\enquote {\bibinfo {title} {{Aharonov-Bohm
  Oscillations in Disordered Topological Insulator Nanowires}},}\ }\href
  {\doibase 10.1103/PhysRevLett.105.156803} {\bibfield  {journal} {\bibinfo
  {journal} {Phys. Rev. Lett.}\ }\textbf {\bibinfo {volume} {105}},\ \bibinfo
  {pages} {156803} (\bibinfo {year} {2010})}\BibitemShut {NoStop}%
\bibitem [{\citenamefont {Zhang}\ and\ \citenamefont
  {Vishwanath}(2010)}]{zhang2010}%
  \BibitemOpen
  \bibfield  {author} {\bibinfo {author} {\bibfnamefont {Yi}~\bibnamefont
  {Zhang}}\ and\ \bibinfo {author} {\bibfnamefont {Ashvin}\ \bibnamefont
  {Vishwanath}},\ }\bibfield  {title} {\enquote {\bibinfo {title} {{Anomalous
  Aharonov-Bohm Conductance Oscillations from Topological Insulator Surface
  States}},}\ }\href {\doibase 10.1103/PhysRevLett.105.206601} {\bibfield
  {journal} {\bibinfo  {journal} {Phys. Rev. Lett.}\ }\textbf {\bibinfo
  {volume} {105}},\ \bibinfo {pages} {206601} (\bibinfo {year}
  {2010})}\BibitemShut {NoStop}%
\bibitem [{\citenamefont {Rosenberg}\ \emph {et~al.}(2010)\citenamefont
  {Rosenberg}, \citenamefont {Guo},\ and\ \citenamefont
  {Franz}}]{rosenberg2010}%
  \BibitemOpen
  \bibfield  {author} {\bibinfo {author} {\bibfnamefont {G.}~\bibnamefont
  {Rosenberg}}, \bibinfo {author} {\bibfnamefont {H.-M.}\ \bibnamefont {Guo}},
  \ and\ \bibinfo {author} {\bibfnamefont {M.}~\bibnamefont {Franz}},\
  }\bibfield  {title} {\enquote {\bibinfo {title} {Wormhole effect in a strong
  topological insulator},}\ }\href {\doibase 10.1103/PhysRevB.82.041104}
  {\bibfield  {journal} {\bibinfo  {journal} {Phys. Rev. B}\ }\textbf {\bibinfo
  {volume} {82}},\ \bibinfo {pages} {041104(R)} (\bibinfo {year}
  {2010})}\BibitemShut {NoStop}%
\bibitem [{\citenamefont {Bardarson}\ and\ \citenamefont
  {Moore}(2013)}]{bardarson2013}%
  \BibitemOpen
  \bibfield  {author} {\bibinfo {author} {\bibfnamefont {Jens~H}\ \bibnamefont
  {Bardarson}}\ and\ \bibinfo {author} {\bibfnamefont {Joel~E}\ \bibnamefont
  {Moore}},\ }\bibfield  {title} {\enquote {\bibinfo {title} {{Quantum
  interference and Aharonov{\textendash}Bohm oscillations in topological
  insulators}},}\ }\href {\doibase 10.1088/0034-4885/76/5/056501} {\bibfield
  {journal} {\bibinfo  {journal} {Reports on Progress in Physics}\ }\textbf
  {\bibinfo {volume} {76}},\ \bibinfo {pages} {056501} (\bibinfo {year}
  {2013})}\BibitemShut {NoStop}%
\bibitem [{\citenamefont {Cho}\ \emph {et~al.}(2015)\citenamefont {Cho},
  \citenamefont {Dellabetta}, \citenamefont {Zhong}, \citenamefont
  {Schneeloch}, \citenamefont {Liu}, \citenamefont {Gu}, \citenamefont
  {Gilbert},\ and\ \citenamefont {Mason}}]{cho2015}%
  \BibitemOpen
  \bibfield  {author} {\bibinfo {author} {\bibfnamefont {Sungjae}\ \bibnamefont
  {Cho}}, \bibinfo {author} {\bibfnamefont {Brian}\ \bibnamefont {Dellabetta}},
  \bibinfo {author} {\bibfnamefont {Ruidan}\ \bibnamefont {Zhong}}, \bibinfo
  {author} {\bibfnamefont {John}\ \bibnamefont {Schneeloch}}, \bibinfo {author}
  {\bibfnamefont {Tiansheng}\ \bibnamefont {Liu}}, \bibinfo {author}
  {\bibfnamefont {Genda}\ \bibnamefont {Gu}}, \bibinfo {author} {\bibfnamefont
  {Matthew~J.}\ \bibnamefont {Gilbert}}, \ and\ \bibinfo {author}
  {\bibfnamefont {Nadya}\ \bibnamefont {Mason}},\ }\bibfield  {title} {\enquote
  {\bibinfo {title} {{Aharonov-Bohm oscillations in a quasi-ballistic
  three-dimensional topological insulator nanowire}},}\ }\href
  {https://doi.org/10.1038/ncomms8634} {\bibfield  {journal} {\bibinfo
  {journal} {Nature Communications}\ }\textbf {\bibinfo {volume} {6}},\
  \bibinfo {pages} {7634} (\bibinfo {year} {2015})}\BibitemShut {NoStop}%
\bibitem [{\citenamefont {Ziegler}\ \emph {et~al.}(2018)\citenamefont
  {Ziegler}, \citenamefont {Kozlovsky}, \citenamefont {Gorini}, \citenamefont
  {Liu}, \citenamefont {Weish\"aupl}, \citenamefont {Maier}, \citenamefont
  {Fischer}, \citenamefont {Kozlov}, \citenamefont {Kvon}, \citenamefont
  {Mikhailov}, \citenamefont {Dvoretsky}, \citenamefont {Richter},\ and\
  \citenamefont {Weiss}}]{ziegler2018}%
  \BibitemOpen
  \bibfield  {author} {\bibinfo {author} {\bibfnamefont {J.}~\bibnamefont
  {Ziegler}}, \bibinfo {author} {\bibfnamefont {R.}~\bibnamefont {Kozlovsky}},
  \bibinfo {author} {\bibfnamefont {C.}~\bibnamefont {Gorini}}, \bibinfo
  {author} {\bibfnamefont {M.-H.}\ \bibnamefont {Liu}}, \bibinfo {author}
  {\bibfnamefont {S.}~\bibnamefont {Weish\"aupl}}, \bibinfo {author}
  {\bibfnamefont {H.}~\bibnamefont {Maier}}, \bibinfo {author} {\bibfnamefont
  {R.}~\bibnamefont {Fischer}}, \bibinfo {author} {\bibfnamefont {D.~A.}\
  \bibnamefont {Kozlov}}, \bibinfo {author} {\bibfnamefont {Z.~D.}\
  \bibnamefont {Kvon}}, \bibinfo {author} {\bibfnamefont {N.}~\bibnamefont
  {Mikhailov}}, \bibinfo {author} {\bibfnamefont {S.~A.}\ \bibnamefont
  {Dvoretsky}}, \bibinfo {author} {\bibfnamefont {K.}~\bibnamefont {Richter}},
  \ and\ \bibinfo {author} {\bibfnamefont {D.}~\bibnamefont {Weiss}},\
  }\bibfield  {title} {\enquote {\bibinfo {title} {{Probing spin helical
  surface states in topological HgTe nanowires}},}\ }\href {\doibase
  10.1103/PhysRevB.97.035157} {\bibfield  {journal} {\bibinfo  {journal} {Phys.
  Rev. B}\ }\textbf {\bibinfo {volume} {97}},\ \bibinfo {pages} {035157}
  (\bibinfo {year} {2018})}\BibitemShut {NoStop}%
\bibitem [{\citenamefont {Imura}\ \emph {et~al.}(2012)\citenamefont {Imura},
  \citenamefont {Yoshimura}, \citenamefont {Takane},\ and\ \citenamefont
  {Fukui}}]{imura2012}%
  \BibitemOpen
  \bibfield  {author} {\bibinfo {author} {\bibfnamefont {Ken-Ichiro}\
  \bibnamefont {Imura}}, \bibinfo {author} {\bibfnamefont {Yukinori}\
  \bibnamefont {Yoshimura}}, \bibinfo {author} {\bibfnamefont {Yositake}\
  \bibnamefont {Takane}}, \ and\ \bibinfo {author} {\bibfnamefont {Takahiro}\
  \bibnamefont {Fukui}},\ }\bibfield  {title} {\enquote {\bibinfo {title}
  {Spherical topological insulator},}\ }\href {\doibase
  10.1103/PhysRevB.86.235119} {\bibfield  {journal} {\bibinfo  {journal} {Phys.
  Rev. B}\ }\textbf {\bibinfo {volume} {86}},\ \bibinfo {pages} {235119}
  (\bibinfo {year} {2012})}\BibitemShut {NoStop}%
\bibitem [{\citenamefont {Takane}\ and\ \citenamefont
  {Imura}(2013)}]{takane2013}%
  \BibitemOpen
  \bibfield  {author} {\bibinfo {author} {\bibfnamefont {Yositake}\
  \bibnamefont {Takane}}\ and\ \bibinfo {author} {\bibfnamefont {Ken-Ichiro}\
  \bibnamefont {Imura}},\ }\bibfield  {title} {\enquote {\bibinfo {title}
  {{Unified Description of Dirac Electrons on a Curved Surface of Topological
  Insulators}},}\ }\href {\doibase 10.7566/JPSJ.82.074712} {\bibfield
  {journal} {\bibinfo  {journal} {Journal of the Physical Society of Japan}\
  }\textbf {\bibinfo {volume} {82}},\ \bibinfo {pages} {074712} (\bibinfo
  {year} {2013})}\BibitemShut {NoStop}%
\bibitem [{\citenamefont {Xypakis}\ \emph {et~al.}(2020)\citenamefont
  {Xypakis}, \citenamefont {Rhim}, \citenamefont {Bardarson},\ and\
  \citenamefont {Ilan}}]{xypakis2017}%
  \BibitemOpen
  \bibfield  {author} {\bibinfo {author} {\bibfnamefont {Emmanouil}\
  \bibnamefont {Xypakis}}, \bibinfo {author} {\bibfnamefont {Jun-Won}\
  \bibnamefont {Rhim}}, \bibinfo {author} {\bibfnamefont {Jens~H.}\
  \bibnamefont {Bardarson}}, \ and\ \bibinfo {author} {\bibfnamefont {Roni}\
  \bibnamefont {Ilan}},\ }\bibfield  {title} {\enquote {\bibinfo {title}
  {{Perfect transmission and Aharanov-Bohm oscillations in topological
  insulator nanowires with nonuniform cross section}},}\ }\href {\doibase
  10.1103/PhysRevB.101.045401} {\bibfield  {journal} {\bibinfo  {journal}
  {Phys. Rev. B}\ }\textbf {\bibinfo {volume} {101}},\ \bibinfo {pages}
  {045401} (\bibinfo {year} {2020})}\BibitemShut {NoStop}%
\bibitem [{\citenamefont {Zhang}\ \emph
  {et~al.}(2011{\natexlab{a}})\citenamefont {Zhang}, \citenamefont {Wang},\
  and\ \citenamefont {Xie}}]{zhang2011}%
  \BibitemOpen
  \bibfield  {author} {\bibinfo {author} {\bibfnamefont {Yan-Yang}\
  \bibnamefont {Zhang}}, \bibinfo {author} {\bibfnamefont {Xiang-Rong}\
  \bibnamefont {Wang}}, \ and\ \bibinfo {author} {\bibfnamefont {X~C}\
  \bibnamefont {Xie}},\ }\bibfield  {title} {\enquote {\bibinfo {title}
  {{Three-dimensional topological insulator in a magnetic field: chiral side
  surface states and quantized Hall conductance}},}\ }\href {\doibase
  10.1088/0953-8984/24/1/015004} {\bibfield  {journal} {\bibinfo  {journal}
  {Journal of Physics: Condensed Matter}\ }\textbf {\bibinfo {volume} {24}},\
  \bibinfo {pages} {015004} (\bibinfo {year} {2011}{\natexlab{a}})}\BibitemShut
  {NoStop}%
\bibitem [{\citenamefont {Vafek}(2011)}]{vafek2011}%
  \BibitemOpen
  \bibfield  {author} {\bibinfo {author} {\bibfnamefont {Oskar}\ \bibnamefont
  {Vafek}},\ }\bibfield  {title} {\enquote {\bibinfo {title} {{Quantum Hall
  effect in a singly and doubly connected three-dimensional topological
  insulator}},}\ }\href {\doibase 10.1103/PhysRevB.84.245417} {\bibfield
  {journal} {\bibinfo  {journal} {Phys. Rev. B}\ }\textbf {\bibinfo {volume}
  {84}},\ \bibinfo {pages} {245417} (\bibinfo {year} {2011})}\BibitemShut
  {NoStop}%
\bibitem [{\citenamefont {Sitte}\ \emph {et~al.}(2012)\citenamefont {Sitte},
  \citenamefont {Rosch}, \citenamefont {Altman},\ and\ \citenamefont
  {Fritz}}]{sitte2012}%
  \BibitemOpen
  \bibfield  {author} {\bibinfo {author} {\bibfnamefont {M.}~\bibnamefont
  {Sitte}}, \bibinfo {author} {\bibfnamefont {A.}~\bibnamefont {Rosch}},
  \bibinfo {author} {\bibfnamefont {E.}~\bibnamefont {Altman}}, \ and\ \bibinfo
  {author} {\bibfnamefont {L.}~\bibnamefont {Fritz}},\ }\bibfield  {title}
  {\enquote {\bibinfo {title} {{Topological Insulators in Magnetic Fields:
  Quantum Hall Effect and Edge Channels with a Nonquantized
  $\ensuremath{\theta}$ Term}},}\ }\href {\doibase
  10.1103/PhysRevLett.108.126807} {\bibfield  {journal} {\bibinfo  {journal}
  {Phys. Rev. Lett.}\ }\textbf {\bibinfo {volume} {108}},\ \bibinfo {pages}
  {126807} (\bibinfo {year} {2012})}\BibitemShut {NoStop}%
\bibitem [{\citenamefont {Brey}\ and\ \citenamefont {Fertig}(2014)}]{brey2014}%
  \BibitemOpen
  \bibfield  {author} {\bibinfo {author} {\bibfnamefont {L.}~\bibnamefont
  {Brey}}\ and\ \bibinfo {author} {\bibfnamefont {H.~A.}\ \bibnamefont
  {Fertig}},\ }\bibfield  {title} {\enquote {\bibinfo {title} {Electronic
  states of wires and slabs of topological insulators: Quantum hall effects and
  edge transport},}\ }\href {\doibase 10.1103/PhysRevB.89.085305} {\bibfield
  {journal} {\bibinfo  {journal} {Phys. Rev. B}\ }\textbf {\bibinfo {volume}
  {89}},\ \bibinfo {pages} {085305} (\bibinfo {year} {2014})}\BibitemShut
  {NoStop}%
\bibitem [{\citenamefont {K\"onig}\ \emph {et~al.}(2014)\citenamefont
  {K\"onig}, \citenamefont {Ostrovsky}, \citenamefont {Protopopov},
  \citenamefont {Gornyi}, \citenamefont {Burmistrov},\ and\ \citenamefont
  {Mirlin}}]{koenig2014}%
  \BibitemOpen
  \bibfield  {author} {\bibinfo {author} {\bibfnamefont {E.~J.}\ \bibnamefont
  {K\"onig}}, \bibinfo {author} {\bibfnamefont {P.~M.}\ \bibnamefont
  {Ostrovsky}}, \bibinfo {author} {\bibfnamefont {I.~V.}\ \bibnamefont
  {Protopopov}}, \bibinfo {author} {\bibfnamefont {I.~V.}\ \bibnamefont
  {Gornyi}}, \bibinfo {author} {\bibfnamefont {I.~S.}\ \bibnamefont
  {Burmistrov}}, \ and\ \bibinfo {author} {\bibfnamefont {A.~D.}\ \bibnamefont
  {Mirlin}},\ }\bibfield  {title} {\enquote {\bibinfo {title} {{Half-integer
  quantum Hall effect of disordered Dirac fermions at a topological insulator
  surface}},}\ }\href {\doibase 10.1103/PhysRevB.90.165435} {\bibfield
  {journal} {\bibinfo  {journal} {Phys. Rev. B}\ }\textbf {\bibinfo {volume}
  {90}},\ \bibinfo {pages} {165435} (\bibinfo {year} {2014})}\BibitemShut
  {NoStop}%
\bibitem [{\citenamefont {Kessel}\ \emph {et~al.}(2017)\citenamefont {Kessel},
  \citenamefont {Hajer}, \citenamefont {Karczewski}, \citenamefont
  {Schumacher}, \citenamefont {Br{\"u}ne}, \citenamefont {Buhmann},\ and\
  \citenamefont {Molenkamp}}]{Kessel_2017}%
  \BibitemOpen
  \bibfield  {author} {\bibinfo {author} {\bibfnamefont {M.}~\bibnamefont
  {Kessel}}, \bibinfo {author} {\bibfnamefont {J.}~\bibnamefont {Hajer}},
  \bibinfo {author} {\bibfnamefont {G.}~\bibnamefont {Karczewski}}, \bibinfo
  {author} {\bibfnamefont {C.}~\bibnamefont {Schumacher}}, \bibinfo {author}
  {\bibfnamefont {C.}~\bibnamefont {Br{\"u}ne}}, \bibinfo {author}
  {\bibfnamefont {H.}~\bibnamefont {Buhmann}}, \ and\ \bibinfo {author}
  {\bibfnamefont {L.~W.}\ \bibnamefont {Molenkamp}},\ }\bibfield  {title}
  {\enquote {\bibinfo {title} {{CdTe-HgTe core-shell nanowire growth controlled
  by RHEED}},}\ }\href {\doibase 10.1103/PhysRevMaterials.1.023401} {\bibfield
  {journal} {\bibinfo  {journal} {Phys. Rev. Materials}\ }\textbf {\bibinfo
  {volume} {1}},\ \bibinfo {pages} {023401} (\bibinfo {year}
  {2017})}\BibitemShut {NoStop}%
\bibitem [{\citenamefont {Kong}\ \emph {et~al.}(2011)\citenamefont {Kong},
  \citenamefont {Chen}, \citenamefont {Cha}, \citenamefont {Zhang},
  \citenamefont {Analytis}, \citenamefont {Lai}, \citenamefont {Liu},
  \citenamefont {Hong}, \citenamefont {Koski}, \citenamefont {Mo},
  \citenamefont {Hussain}, \citenamefont {Fisher}, \citenamefont {Shen},\ and\
  \citenamefont {Cui}}]{Kong_2011}%
  \BibitemOpen
  \bibfield  {author} {\bibinfo {author} {\bibfnamefont {Desheng}\ \bibnamefont
  {Kong}}, \bibinfo {author} {\bibfnamefont {Yulin}\ \bibnamefont {Chen}},
  \bibinfo {author} {\bibfnamefont {Judy~J.}\ \bibnamefont {Cha}}, \bibinfo
  {author} {\bibfnamefont {Qianfan}\ \bibnamefont {Zhang}}, \bibinfo {author}
  {\bibfnamefont {James~G.}\ \bibnamefont {Analytis}}, \bibinfo {author}
  {\bibfnamefont {Keji}\ \bibnamefont {Lai}}, \bibinfo {author} {\bibfnamefont
  {Zhongkai}\ \bibnamefont {Liu}}, \bibinfo {author} {\bibfnamefont
  {Seung~Sae}\ \bibnamefont {Hong}}, \bibinfo {author} {\bibfnamefont
  {Kristie~J.}\ \bibnamefont {Koski}}, \bibinfo {author} {\bibfnamefont
  {Sung-Kwan}\ \bibnamefont {Mo}}, \bibinfo {author} {\bibfnamefont {Zahid}\
  \bibnamefont {Hussain}}, \bibinfo {author} {\bibfnamefont {Ian~R.}\
  \bibnamefont {Fisher}}, \bibinfo {author} {\bibfnamefont {Zhi-Xun}\
  \bibnamefont {Shen}}, \ and\ \bibinfo {author} {\bibfnamefont
  {Yi}~\bibnamefont {Cui}},\ }\bibfield  {title} {\enquote {\bibinfo {title}
  {{Ambipolar field effect in the ternary topological insulator
  $({\mathrm{Bi}}_{x}{\mathrm{Sb}}_{1-x})_2{\mathrm{Te}}_3$ by composition
  tuning}},}\ }\href {https://doi.org/10.1038/nnano.2011.172} {\bibfield
  {journal} {\bibinfo  {journal} {Nature Nanotechnology}\ }\textbf {\bibinfo
  {volume} {6}},\ \bibinfo {pages} {705--709} (\bibinfo {year}
  {2011})}\BibitemShut {NoStop}%
\bibitem [{\citenamefont {Zhang}\ \emph
  {et~al.}(2011{\natexlab{b}})\citenamefont {Zhang}, \citenamefont {Chang},
  \citenamefont {Zhang}, \citenamefont {Wen}, \citenamefont {Feng},
  \citenamefont {Li}, \citenamefont {Liu}, \citenamefont {He}, \citenamefont
  {Wang}, \citenamefont {Chen}, \citenamefont {Xue}, \citenamefont {Ma},\ and\
  \citenamefont {Wang}}]{ZhangJ_2011}%
  \BibitemOpen
  \bibfield  {author} {\bibinfo {author} {\bibfnamefont {Jinsong}\ \bibnamefont
  {Zhang}}, \bibinfo {author} {\bibfnamefont {Cui-Zu}\ \bibnamefont {Chang}},
  \bibinfo {author} {\bibfnamefont {Zuocheng}\ \bibnamefont {Zhang}}, \bibinfo
  {author} {\bibfnamefont {Jing}\ \bibnamefont {Wen}}, \bibinfo {author}
  {\bibfnamefont {Xiao}\ \bibnamefont {Feng}}, \bibinfo {author} {\bibfnamefont
  {Kang}\ \bibnamefont {Li}}, \bibinfo {author} {\bibfnamefont {Minhao}\
  \bibnamefont {Liu}}, \bibinfo {author} {\bibfnamefont {Ke}~\bibnamefont
  {He}}, \bibinfo {author} {\bibfnamefont {Lili}\ \bibnamefont {Wang}},
  \bibinfo {author} {\bibfnamefont {Xi}~\bibnamefont {Chen}}, \bibinfo {author}
  {\bibfnamefont {Qi-Kun}\ \bibnamefont {Xue}}, \bibinfo {author}
  {\bibfnamefont {Xucun}\ \bibnamefont {Ma}}, \ and\ \bibinfo {author}
  {\bibfnamefont {Yayu}\ \bibnamefont {Wang}},\ }\bibfield  {title} {\enquote
  {\bibinfo {title} {{Band structure engineering in
  $({\mathrm{Bi}}_{1-x}{\mathrm{Sb}}_{x})_2{\mathrm{Te}}_3$ ternary topological
  insulators}},}\ }\href {https://doi.org/10.1038/ncomms1588} {\bibfield
  {journal} {\bibinfo  {journal} {Nature Communications}\ }\textbf {\bibinfo
  {volume} {2}},\ \bibinfo {pages} {574} (\bibinfo {year}
  {2011}{\natexlab{b}})}\BibitemShut {NoStop}%
\bibitem [{\citenamefont {Zhou}\ \emph {et~al.}(2012)\citenamefont {Zhou},
  \citenamefont {Liu}, \citenamefont {Analytis}, \citenamefont {Igarashi},
  \citenamefont {Mo}, \citenamefont {Lu}, \citenamefont {Moore}, \citenamefont
  {Fisher}, \citenamefont {Sasagawa}, \citenamefont {Shen}, \citenamefont
  {Hussain},\ and\ \citenamefont {Chen}}]{Zhou_2012}%
  \BibitemOpen
  \bibfield  {author} {\bibinfo {author} {\bibfnamefont {Bo}~\bibnamefont
  {Zhou}}, \bibinfo {author} {\bibfnamefont {Z~K}\ \bibnamefont {Liu}},
  \bibinfo {author} {\bibfnamefont {J~G}\ \bibnamefont {Analytis}}, \bibinfo
  {author} {\bibfnamefont {K}~\bibnamefont {Igarashi}}, \bibinfo {author}
  {\bibfnamefont {S~K}\ \bibnamefont {Mo}}, \bibinfo {author} {\bibfnamefont
  {D~H}\ \bibnamefont {Lu}}, \bibinfo {author} {\bibfnamefont {R~G}\
  \bibnamefont {Moore}}, \bibinfo {author} {\bibfnamefont {I~R}\ \bibnamefont
  {Fisher}}, \bibinfo {author} {\bibfnamefont {T}~\bibnamefont {Sasagawa}},
  \bibinfo {author} {\bibfnamefont {Z~X}\ \bibnamefont {Shen}}, \bibinfo
  {author} {\bibfnamefont {Z}~\bibnamefont {Hussain}}, \ and\ \bibinfo {author}
  {\bibfnamefont {Y~L}\ \bibnamefont {Chen}},\ }\bibfield  {title} {\enquote
  {\bibinfo {title} {Controlling the carriers of topological insulators by bulk
  and surface doping},}\ }\href {\doibase 10.1088/0268-1242/27/12/124002}
  {\bibfield  {journal} {\bibinfo  {journal} {Semiconductor Science and
  Technology}\ }\textbf {\bibinfo {volume} {27}},\ \bibinfo {pages} {124002}
  (\bibinfo {year} {2012})}\BibitemShut {NoStop}%
\bibitem [{\citenamefont {Fecko}(2006)}]{fecko2006}%
  \BibitemOpen
  \bibfield  {author} {\bibinfo {author} {\bibfnamefont {Marián}\ \bibnamefont
  {Fecko}},\ }\enquote {\bibinfo {title} {{Spinor fields and the Dirac
  operator}},}\ in\ \href {\doibase 10.1017/CBO9780511755590.024} {\emph
  {\bibinfo {booktitle} {Differential Geometry and Lie Groups for
  Physicists}}}\ (\bibinfo  {publisher} {Cambridge University Press},\ \bibinfo
  {year} {2006})\ pp.\ \bibinfo {pages} {635--672}\BibitemShut {NoStop}%
\bibitem [{\citenamefont {Koke}\ \emph {et~al.}(2016)\citenamefont {Koke},
  \citenamefont {Noh},\ and\ \citenamefont {Angelakis}}]{koke2016}%
  \BibitemOpen
  \bibfield  {author} {\bibinfo {author} {\bibfnamefont {Christian}\
  \bibnamefont {Koke}}, \bibinfo {author} {\bibfnamefont {Changsuk}\
  \bibnamefont {Noh}}, \ and\ \bibinfo {author} {\bibfnamefont {Dimitris~G.}\
  \bibnamefont {Angelakis}},\ }\bibfield  {title} {\enquote {\bibinfo {title}
  {Dirac equation in 2-dimensional curved spacetime, particle creation, and
  coupled waveguide arrays},}\ }\href {\doibase
  https://doi.org/10.1016/j.aop.2016.08.013} {\bibfield  {journal} {\bibinfo
  {journal} {Annals of Physics}\ }\textbf {\bibinfo {volume} {374}},\ \bibinfo
  {pages} {162 -- 178} (\bibinfo {year} {2016})}\BibitemShut {NoStop}%
\bibitem [{\citenamefont {Allain}\ and\ \citenamefont
  {Fuchs}(2011)}]{allain2011}%
  \BibitemOpen
  \bibfield  {author} {\bibinfo {author} {\bibfnamefont {P.~E.}\ \bibnamefont
  {Allain}}\ and\ \bibinfo {author} {\bibfnamefont {J.~N.}\ \bibnamefont
  {Fuchs}},\ }\bibfield  {title} {\enquote {\bibinfo {title} {Klein tunneling
  in graphene: optics with massless electrons},}\ }\href {\doibase
  10.1140/epjb/e2011-20351-3} {\bibfield  {journal} {\bibinfo  {journal} {The
  European Physical Journal B}\ }\textbf {\bibinfo {volume} {83}},\ \bibinfo
  {pages} {301} (\bibinfo {year} {2011})}\BibitemShut {NoStop}%
\bibitem [{\citenamefont {De~Martino}\ \emph {et~al.}(2007)\citenamefont
  {De~Martino}, \citenamefont {Dell'Anna},\ and\ \citenamefont
  {Egger}}]{demartino2007}%
  \BibitemOpen
  \bibfield  {author} {\bibinfo {author} {\bibfnamefont {A.}~\bibnamefont
  {De~Martino}}, \bibinfo {author} {\bibfnamefont {L.}~\bibnamefont
  {Dell'Anna}}, \ and\ \bibinfo {author} {\bibfnamefont {R.}~\bibnamefont
  {Egger}},\ }\bibfield  {title} {\enquote {\bibinfo {title} {{Magnetic
  Confinement of Massless Dirac Fermions in Graphene}},}\ }\href {\doibase
  10.1103/PhysRevLett.98.066802} {\bibfield  {journal} {\bibinfo  {journal}
  {Phys. Rev. Lett.}\ }\textbf {\bibinfo {volume} {98}},\ \bibinfo {pages}
  {066802} (\bibinfo {year} {2007})}\BibitemShut {NoStop}%
\bibitem [{\citenamefont {Ramezani~Masir}\ \emph {et~al.}(2008)\citenamefont
  {Ramezani~Masir}, \citenamefont {Vasilopoulos}, \citenamefont {Matulis},\
  and\ \citenamefont {Peeters}}]{masir2008}%
  \BibitemOpen
  \bibfield  {author} {\bibinfo {author} {\bibfnamefont {M.}~\bibnamefont
  {Ramezani~Masir}}, \bibinfo {author} {\bibfnamefont {P.}~\bibnamefont
  {Vasilopoulos}}, \bibinfo {author} {\bibfnamefont {A.}~\bibnamefont
  {Matulis}}, \ and\ \bibinfo {author} {\bibfnamefont {F.~M.}\ \bibnamefont
  {Peeters}},\ }\bibfield  {title} {\enquote {\bibinfo {title}
  {Direction-dependent tunneling through nanostructured magnetic barriers in
  graphene},}\ }\href {\doibase 10.1103/PhysRevB.77.235443} {\bibfield
  {journal} {\bibinfo  {journal} {Phys. Rev. B}\ }\textbf {\bibinfo {volume}
  {77}},\ \bibinfo {pages} {235443} (\bibinfo {year} {2008})}\BibitemShut
  {NoStop}%
\bibitem [{\citenamefont {Groth}\ \emph {et~al.}(2014)\citenamefont {Groth},
  \citenamefont {Wimmer}, \citenamefont {Akhmerov},\ and\ \citenamefont
  {Waintal}}]{groth2014}%
  \BibitemOpen
  \bibfield  {author} {\bibinfo {author} {\bibfnamefont {Christoph~W}\
  \bibnamefont {Groth}}, \bibinfo {author} {\bibfnamefont {Michael}\
  \bibnamefont {Wimmer}}, \bibinfo {author} {\bibfnamefont {Anton~R}\
  \bibnamefont {Akhmerov}}, \ and\ \bibinfo {author} {\bibfnamefont {Xavier}\
  \bibnamefont {Waintal}},\ }\bibfield  {title} {\enquote {\bibinfo {title}
  {Kwant: a software package for quantum transport},}\ }\href {\doibase
  10.1088/1367-2630/16/6/063065} {\bibfield  {journal} {\bibinfo  {journal}
  {New Journal of Physics}\ }\textbf {\bibinfo {volume} {16}},\ \bibinfo
  {pages} {063065} (\bibinfo {year} {2014})}\BibitemShut {NoStop}%
\bibitem [{Note2()}]{Note2}%
  \BibitemOpen
  \bibinfo {note} {As can be seen from Fig.~\ref {fig:nanocone}, the twofold
  angular momentum degeneracy gets lifted for $B\not =0$ and is never restored
  for higher $B$, due to the out-of-surface component. This is in marked
  contrast to cylindrical TINWs.}\BibitemShut {Stop}%
\bibitem [{\citenamefont {Cho}\ \emph {et~al.}(2012)\citenamefont {Cho},
  \citenamefont {Kim}, \citenamefont {Syers}, \citenamefont {Butch},
  \citenamefont {Paglione},\ and\ \citenamefont {Fuhrer}}]{cho2012}%
  \BibitemOpen
  \bibfield  {author} {\bibinfo {author} {\bibfnamefont {Sungjae}\ \bibnamefont
  {Cho}}, \bibinfo {author} {\bibfnamefont {Dohun}\ \bibnamefont {Kim}},
  \bibinfo {author} {\bibfnamefont {Paul}\ \bibnamefont {Syers}}, \bibinfo
  {author} {\bibfnamefont {Nicholas~P.}\ \bibnamefont {Butch}}, \bibinfo
  {author} {\bibfnamefont {Johnpierre}\ \bibnamefont {Paglione}}, \ and\
  \bibinfo {author} {\bibfnamefont {Michael~S.}\ \bibnamefont {Fuhrer}},\
  }\bibfield  {title} {\enquote {\bibinfo {title} {{Topological Insulator
  Quantum Dot with Tunable Barriers}},}\ }\href {\doibase 10.1021/nl203851g}
  {\bibfield  {journal} {\bibinfo  {journal} {Nano Lett.}\ }\textbf {\bibinfo
  {volume} {12}},\ \bibinfo {pages} {469--472} (\bibinfo {year}
  {2012})}\BibitemShut {NoStop}%
\bibitem [{\citenamefont {Beenakker}(1991)}]{beenakker1991}%
  \BibitemOpen
  \bibfield  {author} {\bibinfo {author} {\bibfnamefont {C.~W.~J.}\
  \bibnamefont {Beenakker}},\ }\bibfield  {title} {\enquote {\bibinfo {title}
  {{Theory of Coulomb-blockade oscillations in the conductance of a quantum
  dot}},}\ }\href {\doibase 10.1103/PhysRevB.44.1646} {\bibfield  {journal}
  {\bibinfo  {journal} {Phys. Rev. B}\ }\textbf {\bibinfo {volume} {44}},\
  \bibinfo {pages} {1646--1656} (\bibinfo {year} {1991})}\BibitemShut {NoStop}%
\bibitem [{\citenamefont {Ihn}(2010)}]{Ihn}%
  \BibitemOpen
  \bibfield  {author} {\bibinfo {author} {\bibfnamefont {T.}~\bibnamefont
  {Ihn}},\ }\href {https://books.google.fr/books?id=PAtqPgAACAAJ} {\emph
  {\bibinfo {title} {Semiconductor Nanostructures: Quantum States and
  Electronic Transport}}}\ (\bibinfo  {publisher} {OUP Oxford},\ \bibinfo
  {year} {2010})\BibitemShut {NoStop}%
\bibitem [{\citenamefont {Mu\~{n}oz Rojo}\ \emph {et~al.}(2016)\citenamefont
  {Mu\~{n}oz Rojo}, \citenamefont {Zhang}, \citenamefont {Manzano},
  \citenamefont {Alvaro}, \citenamefont {Gooth}, \citenamefont {Salmeron},\
  and\ \citenamefont {Martin-Gonzalez}}]{Munoz_2016}%
  \BibitemOpen
  \bibfield  {author} {\bibinfo {author} {\bibfnamefont {Miguel}\ \bibnamefont
  {Mu\~{n}oz Rojo}}, \bibinfo {author} {\bibfnamefont {Yingjie}\ \bibnamefont
  {Zhang}}, \bibinfo {author} {\bibfnamefont {Cristina~V.}\ \bibnamefont
  {Manzano}}, \bibinfo {author} {\bibfnamefont {Raquel}\ \bibnamefont
  {Alvaro}}, \bibinfo {author} {\bibfnamefont {Johannes}\ \bibnamefont
  {Gooth}}, \bibinfo {author} {\bibfnamefont {Miquel}\ \bibnamefont
  {Salmeron}}, \ and\ \bibinfo {author} {\bibfnamefont {Marisol}\ \bibnamefont
  {Martin-Gonzalez}},\ }\bibfield  {title} {\enquote {\bibinfo {title}
  {{Spatial potential ripples of azimuthal surface modes in topological
  insulator ${\mathrm{Bi}}_{2}{\mathrm{Te}}_{3}$ nanowires}},}\ }\href
  {https://doi.org/10.1038/srep19014} {\bibfield  {journal} {\bibinfo
  {journal} {Scientific Reports}\ }\textbf {\bibinfo {volume} {6}},\ \bibinfo
  {pages} {19014} (\bibinfo {year} {2016})}\BibitemShut {NoStop}%
\bibitem [{\citenamefont {Storm}\ \emph {et~al.}(2012)\citenamefont {Storm},
  \citenamefont {Nylund}, \citenamefont {Samuelson},\ and\ \citenamefont
  {Micolich}}]{Storm2012}%
  \BibitemOpen
  \bibfield  {author} {\bibinfo {author} {\bibfnamefont {Kristian}\
  \bibnamefont {Storm}}, \bibinfo {author} {\bibfnamefont {Gustav}\
  \bibnamefont {Nylund}}, \bibinfo {author} {\bibfnamefont {Lars}\ \bibnamefont
  {Samuelson}}, \ and\ \bibinfo {author} {\bibfnamefont {Adam~P.}\ \bibnamefont
  {Micolich}},\ }\bibfield  {title} {\enquote {\bibinfo {title} {{Realizing
  Lateral Wrap-Gated Nanowire FETs: Controlling Gate Length with Chemistry
  Rather than Lithography}},}\ }\href {\doibase 10.1021/nl104403g} {\bibfield
  {journal} {\bibinfo  {journal} {Nano Lett.}\ }\textbf {\bibinfo {volume}
  {12}},\ \bibinfo {pages} {1--6} (\bibinfo {year} {2012})}\BibitemShut
  {NoStop}%
\bibitem [{\citenamefont {Royo}\ \emph {et~al.}(2017)\citenamefont {Royo},
  \citenamefont {De~Luca}, \citenamefont {Rurali},\ and\ \citenamefont
  {Zardo}}]{Royo2017}%
  \BibitemOpen
  \bibfield  {author} {\bibinfo {author} {\bibfnamefont {Miquel}\ \bibnamefont
  {Royo}}, \bibinfo {author} {\bibfnamefont {Marta}\ \bibnamefont {De~Luca}},
  \bibinfo {author} {\bibfnamefont {Riccardo}\ \bibnamefont {Rurali}}, \ and\
  \bibinfo {author} {\bibfnamefont {Ilaria}\ \bibnamefont {Zardo}},\ }\bibfield
   {title} {\enquote {\bibinfo {title} {{A review on III-V core-multishell
  nanowires: growth, properties, and applications}},}\ }\href
  {http://dx.doi.org/10.1088/1361-6463/aa5d8e} {\bibfield  {journal} {\bibinfo
  {journal} {Journal of Physics D: Applied Physics}\ }\textbf {\bibinfo
  {volume} {50}},\ \bibinfo {pages} {143001} (\bibinfo {year}
  {2017})}\BibitemShut {NoStop}%
\bibitem [{\citenamefont {Zierenberg}\ \emph {et~al.}(2017)\citenamefont
  {Zierenberg}, \citenamefont {Fricke}, \citenamefont {Marenz}, \citenamefont
  {Spitzner}, \citenamefont {Blavatska},\ and\ \citenamefont
  {Janke}}]{Zierenberg2017}%
  \BibitemOpen
  \bibfield  {author} {\bibinfo {author} {\bibfnamefont {Johannes}\
  \bibnamefont {Zierenberg}}, \bibinfo {author} {\bibfnamefont {Niklas}\
  \bibnamefont {Fricke}}, \bibinfo {author} {\bibfnamefont {Martin}\
  \bibnamefont {Marenz}}, \bibinfo {author} {\bibfnamefont {F.~P.}\
  \bibnamefont {Spitzner}}, \bibinfo {author} {\bibfnamefont {Viktoria}\
  \bibnamefont {Blavatska}}, \ and\ \bibinfo {author} {\bibfnamefont
  {Wolfhard}\ \bibnamefont {Janke}},\ }\bibfield  {title} {\enquote {\bibinfo
  {title} {Percolation thresholds and fractal dimensions for square and cubic
  lattices with long-range correlated defects},}\ }\href {\doibase
  10.1103/PhysRevE.96.062125} {\bibfield  {journal} {\bibinfo  {journal} {Phys.
  Rev. E}\ }\textbf {\bibinfo {volume} {96}},\ \bibinfo {pages} {062125}
  (\bibinfo {year} {2017})}\BibitemShut {NoStop}%
\bibitem [{\citenamefont {Ghosh}(2008)}]{ghosh2008}%
  \BibitemOpen
  \bibfield  {author} {\bibinfo {author} {\bibfnamefont {Tarun~Kanti}\
  \bibnamefont {Ghosh}},\ }\bibfield  {title} {\enquote {\bibinfo {title}
  {{Exact solutions for a Dirac electron in an exponentially decaying magnetic
  field}},}\ }\href {\doibase 10.1088/0953-8984/21/4/045505} {\bibfield
  {journal} {\bibinfo  {journal} {Journal of Physics: Condensed Matter}\
  }\textbf {\bibinfo {volume} {21}},\ \bibinfo {pages} {045505} (\bibinfo
  {year} {2008})}\BibitemShut {NoStop}%
\bibitem [{\citenamefont {Roy}\ \emph {et~al.}(2012)\citenamefont {Roy},
  \citenamefont {Ghosh},\ and\ \citenamefont {Bhattacharya}}]{roy2012}%
  \BibitemOpen
  \bibfield  {author} {\bibinfo {author} {\bibfnamefont {Pratim}\ \bibnamefont
  {Roy}}, \bibinfo {author} {\bibfnamefont {Tarun~Kanti}\ \bibnamefont
  {Ghosh}}, \ and\ \bibinfo {author} {\bibfnamefont {Kaushik}\ \bibnamefont
  {Bhattacharya}},\ }\bibfield  {title} {\enquote {\bibinfo {title}
  {{Localization of Dirac-like excitations in graphene in the presence of
  smooth inhomogeneous magnetic fields}},}\ }\href {\doibase
  10.1088/0953-8984/24/5/055301} {\bibfield  {journal} {\bibinfo  {journal}
  {Journal of Physics: Condensed Matter}\ }\textbf {\bibinfo {volume} {24}},\
  \bibinfo {pages} {055301} (\bibinfo {year} {2012})}\BibitemShut {NoStop}%
\bibitem [{\citenamefont {Yokoyama}\ \emph {et~al.}(2010)\citenamefont
  {Yokoyama}, \citenamefont {Tanaka},\ and\ \citenamefont
  {Nagaosa}}]{tokoyama2010}%
  \BibitemOpen
  \bibfield  {author} {\bibinfo {author} {\bibfnamefont {T.}~\bibnamefont
  {Yokoyama}}, \bibinfo {author} {\bibfnamefont {Y.}~\bibnamefont {Tanaka}}, \
  and\ \bibinfo {author} {\bibfnamefont {N.}~\bibnamefont {Nagaosa}},\
  }\bibfield  {title} {\enquote {\bibinfo {title} {Anomalous magnetoresistance
  of a two-dimensional ferromagnet/ferromagnet junction on the surface of a
  topological insulator},}\ }\href
  {https://journals.aps.org/prb/abstract/10.1103/PhysRevB.81.121401} {\bibfield
   {journal} {\bibinfo  {journal} {Phys. Rev. B}\ }\textbf {\bibinfo {volume}
  {81}},\ \bibinfo {pages} {121401(R)} (\bibinfo {year} {2010})}\BibitemShut
  {NoStop}%
\bibitem [{\citenamefont {Schwab}\ \emph {et~al.}(2011)\citenamefont {Schwab},
  \citenamefont {Raimondi},\ and\ \citenamefont {Gorini}}]{schwab2011}%
  \BibitemOpen
  \bibfield  {author} {\bibinfo {author} {\bibfnamefont {P.}~\bibnamefont
  {Schwab}}, \bibinfo {author} {\bibfnamefont {R.}~\bibnamefont {Raimondi}}, \
  and\ \bibinfo {author} {\bibfnamefont {C.}~\bibnamefont {Gorini}},\
  }\bibfield  {title} {\enquote {\bibinfo {title} {Spin-charge locking and
  tunneling into a helical metal},}\ }\href
  {https://iopscience.iop.org/article/10.1209/0295-5075/93/67004/pdf}
  {\bibfield  {journal} {\bibinfo  {journal} {EPL}\ }\textbf {\bibinfo {volume}
  {93}},\ \bibinfo {pages} {67004} (\bibinfo {year} {2011})}\BibitemShut
  {NoStop}%
\bibitem [{\citenamefont {Lee}\ \emph {et~al.}(2015)\citenamefont {Lee},
  \citenamefont {Richardella}, \citenamefont {Hickey}, \citenamefont
  {Mkhoyan},\ and\ \citenamefont {Samarth}}]{lee2015}%
  \BibitemOpen
  \bibfield  {author} {\bibinfo {author} {\bibfnamefont {J.~S.}\ \bibnamefont
  {Lee}}, \bibinfo {author} {\bibfnamefont {A.}~\bibnamefont {Richardella}},
  \bibinfo {author} {\bibfnamefont {D.~R.}\ \bibnamefont {Hickey}}, \bibinfo
  {author} {\bibfnamefont {K.~A.}\ \bibnamefont {Mkhoyan}}, \ and\ \bibinfo
  {author} {\bibfnamefont {N.}~\bibnamefont {Samarth}},\ }\bibfield  {title}
  {\enquote {\bibinfo {title} {Mapping the chemical potential dependence of
  current-induced spin polarization in a topological insulator},}\ }\href
  {https://journals.aps.org/prb/abstract/10.1103/PhysRevB.92.155312} {\bibfield
   {journal} {\bibinfo  {journal} {Phys. Rev. B}\ }\textbf {\bibinfo {volume}
  {92}},\ \bibinfo {pages} {155312} (\bibinfo {year} {2015})}\BibitemShut
  {NoStop}%
\bibitem [{\citenamefont {Culcer}(2012)}]{culcer2012}%
  \BibitemOpen
  \bibfield  {author} {\bibinfo {author} {\bibfnamefont {D.}~\bibnamefont
  {Culcer}},\ }\bibfield  {title} {\enquote {\bibinfo {title} {{Transport in
  three-dimensional topological insulators: Theory and experiment}},}\ }\href
  {https://www.sciencedirect.com/science/article/pii/S1386947711003985?via%3Dihub}
  {\bibfield  {journal} {\bibinfo  {journal} {Physica E}\ }\textbf {\bibinfo
  {volume} {44}},\ \bibinfo {pages} {860} (\bibinfo {year} {2012})}\BibitemShut
  {NoStop}%
\bibitem [{\citenamefont {Ando}\ and\ \citenamefont
  {Shiraishi}(2017)}]{ando2017}%
  \BibitemOpen
  \bibfield  {author} {\bibinfo {author} {\bibfnamefont {Y.}~\bibnamefont
  {Ando}}\ and\ \bibinfo {author} {\bibfnamefont {M.}~\bibnamefont
  {Shiraishi}},\ }\bibfield  {title} {\enquote {\bibinfo {title} {{Spin to
  Charge Interconversion Phenomena in the Interface and Surface States}},}\
  }\href {https://journals.jps.jp/doi/abs/10.7566/JPSJ.86.011001} {\bibfield
  {journal} {\bibinfo  {journal} {J. Phys. Soc. Jpn.}\ }\textbf {\bibinfo
  {volume} {86}},\ \bibinfo {pages} {011001} (\bibinfo {year}
  {2017})}\BibitemShut {NoStop}%
\end{thebibliography}%

\end{document}